\documentclass[journal=jctcce,manuscript=article,
               super=true,layout=twocolumn]{achemso}

\usepackage{amsmath}
\usepackage{amssymb}
\usepackage{braket}
\usepackage{bbold}
\usepackage[version=3]{mhchem}
\usepackage{graphicx}
\usepackage{color}
\usepackage{units}
\usepackage{pifont}
\usepackage[caption=false]{subfig}
\usepackage{multirow}
\usepackage[dvipsnames]{xcolor}
\usepackage{hyperref}
\usepackage[final]{changes}
\usepackage{chemformula}

\setkeys{acs}{articletitle = true}
\setkeys{acs}{chaptertitle = true}

\DeclareMathOperator{\Tr}{Tr}

\newcommand{\op}[1]{\hat {#1}}
\newcommand{\trace}[1]{\mathrm{tr}\left(#1\right)}
\newcommand{\matop}[1]{\mathbf{#1}}
\newcommand{\identity}{\op{\mathbb I}}

\author{William Dawson}
\affiliation{RIKEN Center for Computational Science, Kobe, Japan}
\email{william.dawson@riken.jp}

\author{Stephan Mohr}
\affiliation{Barcelona Supercomputing Center (BSC)}

\author{Laura E. Ratcliff}
\affiliation{Department of Materials, Imperial College London, London SW7 2AZ, United Kingdom}

\author{Takahito Nakajima}
\affiliation{RIKEN Center for Computational Science, Kobe, Japan}

\author{Luigi Genovese}
\affiliation{Univ.\ Grenoble Alpes, INAC-MEM, L\_Sim, F-38000 Grenoble, France}
\affiliation{CEA, INAC-MEM, L\_Sim, F-38000 Grenoble, France}
\email{luigi.genovese@cea.fr}

\title{Complexity Reduction in Density Functional Theory Calculations of Large Systems: System Partitioning and Fragment Embedding}

\begin{document}

\begin{tocentry}
\includegraphics[width=7cm, height=3.5cm]{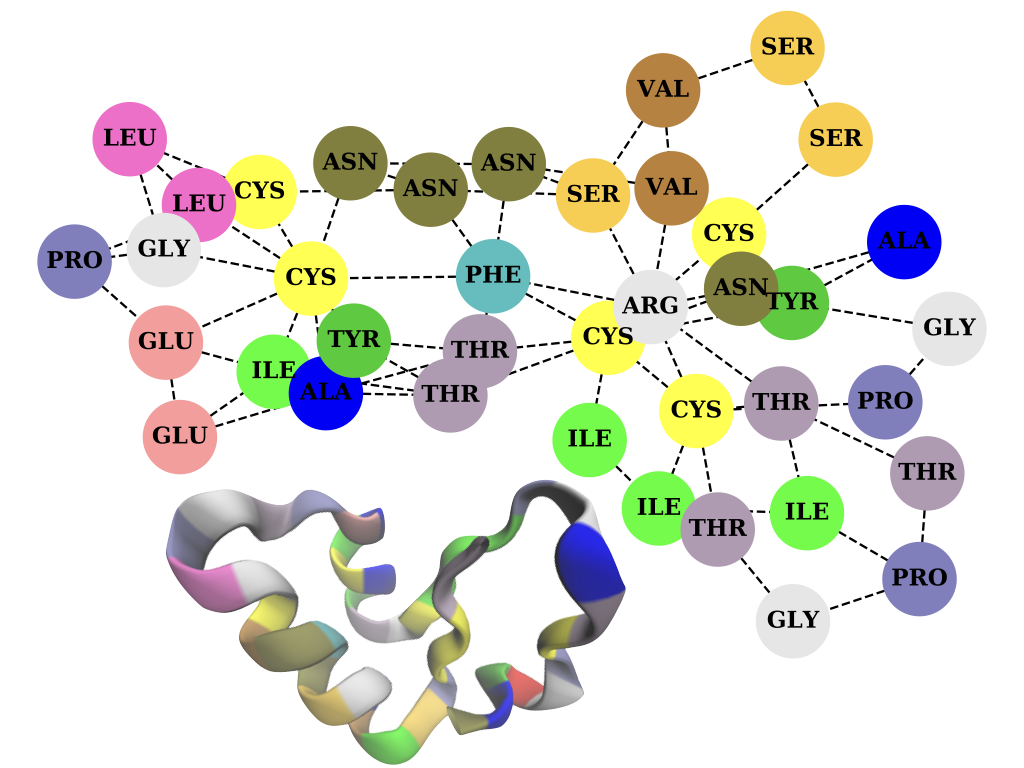}
\end{tocentry}

\begin{abstract}
With the development of low order scaling methods for performing Kohn-Sham
Density Functional Theory, it is now possible to perform fully quantum
mechanical calculations of systems containing tens of thousands of atoms.
However, with an increase in the size of system treated comes an increase
in complexity, making it challenging to analyze such large systems and determine
the cause of emergent properties. To address this issue, in this paper we
present a systematic complexity reduction methodology which can break down
large systems into their constituent fragments, and quantify inter-fragment
interactions. The methodology proposed here requires no a priori information
or user interaction, allowing a single workflow to be automatically applied
to any system of interest. We apply this approach to a variety of
different systems, and show how it allows for the derivation of new system
descriptors, the design of QM/MM partitioning schemes, and the novel
application of graph metrics to molecules and materials.
\end{abstract}

\maketitle

\section{Introduction}
Linear scaling algorithms for Kohn-Sham (KS) Density Functional Theory
(DFT)~\cite{hohenberg-inhomogeneous-1964,kohn-self_consistent-1965}, developed
already some time ago~\cite{goedecker-linear-1999,bowler-O(N)-2012}, have
recently become accessible to a broader community thanks to the introduction
of reliable and robust approaches (see
e.g.\ Ref.~\citenum{ratcliff-2016-challenges} and references therein).
This fact has important consequences for the interpretation and design of
first-principle approaches, as the possibility of tackling systems of
unconventionally large sizes allows for the addressing of new
scientific questions. However, when treating heterogeneous systems, an increase
in system size leads to an increase in complexity, making the interpretation of
computational results challenging.

For a system containing many thousand atoms, it is likely that the fundamental
constituents (or ``moieties'')  of the system are of $\mathcal O(1)$, i.e.\ their
size does not increase with the total number of atoms of the system. It appears
therefore interesting to single out such moieties, and to try to model their
mutual interactions with a less complex description. Thanks to linear
scaling DFT techniques, the full quantum-mechanical (QM) calculation of the
original system can be used as an assessment of the quality of such simplified
descriptions.

When linking together various length scales, such considerations are no longer
optional, but they rather become compulsory. Performing a set of production QM
simulations with an unnecessarily costly approach would result in a study of
poor quality, as the simulation scheme entangles interactions with different
length scales and couplings. In other terms, the dogma ``the more complex the
simulation the better'' is not true in all situations. Taking these considerations
into account allows one to focus on the regions of the system that
\emph{require} a high level of theory, leading to a better understanding of the
fundamental mechanisms and avoiding an unnecessary waste of computational
resources.

In this context --- which we will from now on denote as
``complexity reduction'' --- we briefly want to point out the important difference
between \emph{fragmentation} and \emph{embedding}. In the first case, the full
QM system is partitioned into several fragments, which are each individually
treated at a full QM level, but which are mutually interacting in a simplified
way. Fragmentation methods are conceived to simplify the full ab-initio
calculation of a large QM system, i.e.\ they aim to treat the entire system at
the same level of theory. Famous examples are, for instance, the
Fragment Molecular Orbital approach~\cite{kitaura-fragment-1999,fedorov-extending-2007},
the X-Pol method~\cite{gao-toward-1997,gao-a-molecular-1998,wierzchowski-hydrogen-2003,xie-design-2007,xie-the-variational-2008,xie-incorporation-2008,wang-multilevel-2012,gao-variational-2012},
the Molecular Tailoring Approach~\cite{gadre-molecular-1994,ganesh-molecular-2006,gadre2006molecular,sahu-molecular-2014},
or subsystem DFT~\cite{jacob-subsystem-2014,krishtal-subsystem-2015}.
Embedding methods, on the other hand, aim to split the system into a target region
and an environment, each treated at different computational cost. Embedding approaches
use various levels of theory within a single calculation, thus paving the
way towards coarse grained models which can be used within multi-scale
QM/MM simulations. Among others, we quote here the methods detailed in
Refs.~\citenum{bakowies-hybrid-1996, ESPF_2002, chung2015oniom, lin-QM/MM-2007,
senn-QM/MM-2009, collins2015energy, gordon2011fragmentation,
raghavachari2015accurate, he2014fragment, li2014generalized, hirata2014ab,
sahu2014molecular, pruitt2014efficient, mezey2014fuzzy, herbert2019fantasy}.

A problem common to both fragmentation and embedding methods is how to derive a
general and reliable method for partitioning an arbitrary system into a set of
fragments. As a matter of fact, the concept of fragmentation is to some
extent an ``ad-hoc'' operation, based on the assumption that the system
can be somehow partitioned into subsystems that mutually interact.
In a previous publication~\cite{mohr-fragments1-2017}, we derived a simple
method of determining in a quantitative way whether a chosen fragmentation is
reasonable. If this is the case, the fragments become ``independent'' of each
other and can be assigned ``pseudo-observables'' i.e.\ quantities with an
interpretable physicochemical meaning.

In this paper, we build upon our previous work on evaluating fragments in order
to develop a full methodology for complexity reduction. We will begin in
Sec.~\ref{sec:FaIO} by reintroducing the \textit{purity indicator} as
a measure of fragment quality. Then in Sec.~\ref{sec:SigofPI}, we will
define a new measure called the \textit{fragment bond order}, which
quantifies the interaction strength between fragments. We then will utilize
the fragment bond order to determine the chemical significance of the purity
indicator. In Sec.~\ref{sec:fragandsub} we will further use the
fragment bond order to define an embedding environment for fragments,
and show how that can be used to build a graph like view of a molecular system.
In Sec.~\ref{sec:AF}, we will describe an automatic procedure that uses the
fragment bond order to fragment a given system such that
the purity indicator is close to zero for each fragment. Finally,
we will conclude by demonstrating this methodology on a number of systems,
and discuss how this methodology might bring together the concepts of
fragmentation and embedding, enabling general multilayered schemes for both the
calculation and interpretation of complex, heterogeneous systems.

\section{Fragmentation and Interpretation of Observables}
\label{sec:FaIO}
In a QM system, the expectation value of a one-body observable $\hat O$
can be expressed as $\langle \op O \rangle \equiv \trace{\op F \op O} $, where
we denote by $\op F=\ket{\Psi} \bra{\Psi} = \op F^2 $ the one-body density
matrix of the system, that can be identified in terms of the ground-state
wavefunction $\ket{\Psi}$. When a QM system is susceptible to be genuinely
separable in to fragment states $\ket{\Psi^{\mathcal F}}$, it should be
possible to define a projection operator $\hat W^{\mathcal F}$  associated
with each fragment $\mathcal F$ such that
$\hat W^{\mathcal F}|\Psi\rangle = | \Psi^\mathcal F\rangle $.
Performing such a fragmentation operation \textit{a posteriori} is a procedure
that presents, of course, some degrees of arbitrariness and is susceptible to
provide, in the worst case, a system partitioning into physically meaningless
moieties. The spirit of the fragmentation procedure described
in~\cite{mohr-fragments1-2017} is to provide indicators that helps in assessing
the \emph{physical pertinence} of a given fragmentation. Let us briefly review
this methodology here.

We assume that the density matrix of the system, as well as the projection
operator, can be defined in a set of localized, not necessarily orthonormal,
basis functions $|\phi_\alpha\rangle$ as follows:
\begin{align}
 \hat F &=\sum_{\alpha,\beta} |\phi_\alpha\rangle K_{\alpha\beta}\langle \phi_\beta| \;,
 \label{eq:density_matrix_in_support_function_basis} \\
 \hat W^\mathcal{F} &= \sum_{\mu,\nu}\ket{\phi_\mu}R_{\mu\nu}^\mathcal{F} \bra{\phi_\nu} \;,
 \label{eq:projector_ansatz}
\end{align}
and that a generic one-body operator $\op O$ can be expressed by the matrix
elements $O_{\alpha\beta} = \bra{\phi_\alpha} \op O \ket{\phi_\beta}$. In this
context the overlap matrix
$S_{\alpha\beta}  \equiv \braket{\phi_\alpha | \phi_\beta}$ can be seen as
the matrix representation of the identity operator.

To be meaningful, the fragment projector should satisfy:
\begin{align}
\op W^\mathcal F \op W^\mathcal G = \op W^\mathcal F \delta_{\mathcal{FG}}&\Rightarrow
\matop{R}^\mathcal F \matop S \matop{R}^\mathcal G =
\matop{R}^\mathcal F \delta_{\mathcal{FG}} \label{wortho},\\
\sum_{\mathcal F} \op W^\mathcal F = \identity & \Rightarrow
\sum_{\mathcal F} \matop S \matop{R}^\mathcal F \matop S = \matop S,
\end{align}
which are the obvious orthogonality (including projection) and
resolution-of-the-identity conditions that a reasonable fragmentation should
implement. The latter condition, when combined with the idempotency of
$\op F$, provides $\op F \left(\sum_{\mathcal F}  \op W^\mathcal F\right) \op F = \op F$,
which would imply the interesting equation:
\begin{equation}\label{interesting}
\op W^\mathcal G  \op F \left(\sum_{\mathcal F}  \op W^\mathcal F\right) \op F  \op W^\mathcal G = \op W^\mathcal G  \op F  \op W^\mathcal G.
\end{equation}
Nonetheless, when the system's fragmentation is exact, the fragment density
matrices $\ket{\psi^{\mathcal F}}\bra{\psi^{\mathcal F}}= \op W^\mathcal F \op F \op W^\mathcal F$
should also be idempotent. Together with Eqs.~\eqref{interesting} and \eqref{wortho}
this would imply:
\begin{equation}
\op W^\mathcal G  \op F \left(\sum_{\mathcal F \neq \mathcal G}  \op W^\mathcal F\right) \op F \op W^\mathcal G = 0\;,
\end{equation}
which is a condition that can be realized (excluding pathological situations) by
assuming that \emph{in a meaningful fragmentation, the fragment representation
of the density matrix is block-diagonal}, i.e.
$\op W^\mathcal G  \op F \op W^\mathcal F \equiv \op F W^\mathcal F  \delta_{\mathcal FG} \equiv \op W^\mathcal F  \op F \delta_{\mathcal FG}$ in the span of the basis set chosen.
We may therefore rephrase a meaningful fragmentation as the \emph{purity condition}
$(\op F^{\mathcal F})^2 = \op F^{\mathcal F}$ where we have defined the fragment
density matrix as $ \op F^{\mathcal F} \equiv \op F W^\mathcal F$.
Such a condition depends on the \emph{combination} of the basis set $\phi_\alpha$
and the projection $\matop{R^\mathcal F}$, and cannot be guaranteed a priori,
nor is it a sufficient condition for fragmentation. Simply, when this condition holds,
a system is susceptible to be fragmented by the set of projections identified by
the operators $\op W^\mathcal F$. However, we emphasize that the purity condition
above is more stringent than the idempotency of the operator
$\ket{\psi^{\mathcal F}}\bra{\psi^{\mathcal F}}$ as the latter would impose
the block-like behaviour of $\op F$.

At the same time, an operator can be projected onto the fragment subspace by
defining $\op O^{\mathcal F} \equiv \op W^\mathcal F \op O \op W^\mathcal F$,
which would provide, in the basis set representation,
$\matop{O^{\mathcal{F}}} = \matop{S R^\mathcal{F} O R^\mathcal{F} S}$.

The purity condition is itself represented in the basis set by the expression:
\begin{equation}
\matop{K S} \matop{R^\mathcal{F}} \matop{S K S} \matop{R^\mathcal{F}} = \matop{K S} \matop{R^\mathcal{F}},
\end{equation}
whose trace enables us to introduce the \textit{purity indicator}, defined by:
\begin{equation}
  \Pi_\mathcal F = \frac{1}{q_{\mathcal F}} \Tr\left( \left(\mathbf{ K S}^\mathcal{F} \right)^2 - \mathbf{ K S}^\mathcal{F}\right) \;,
 \label{eq:normalized_purity_indicator}
\end{equation}
where $q_{\mathcal F}$ is the total number of electrons of the isolated
fragment in gas phase
and $\mathbf S^\mathcal{F}\equiv \mathbf S \mathbf R^\mathcal{F} \mathbf S$.
We note that $\Pi \leq 0$ and call \emph{pure} a fragment whose projection
satisfies the condition $\Pi \simeq 0$.

Such a condition, which we emphasize to be \emph{non-linear} in the projector
matrix elements $R^\mathcal{F}$, when fulfilled, enables one to interpret
the fragment-expectation value:
\begin{equation}\label{fragexpval}
\langle \op O \rangle_{\mathcal F} \equiv
\trace{\op F^{\mathcal F} \op O} =
\trace{\matop{ K S R^\mathcal{F} O} }\;\;,
\end{equation}
as a pseudo-observable of the fragment $\mathcal F$. Indeed, by
resolution-of-the-identity, we may decompose the expectation value in to
fragment-wise values, namely
$\langle \op O \rangle = \sum_{\mathcal F} \langle \op O \rangle_{\mathcal F}$.
We retrieve here the \emph{extensivity} of the expectation values: as this condition is
linear in the fragment projection operator, a collection of fragments is itself a
fragment and their expectation value is the sum of the separate contributions.
More importantly, thanks to this property a fragment pseudo-observable can be
\emph{decomposed} into different contributions.
Let $\matop{R^\mathcal F} = \matop{R^\mathcal F_1 + R^\mathcal F_2}$. Even if
the fragments $\mathcal{F}_{1,2}$ were not pure,
still we would have $\langle \op O \rangle_{\mathcal F} = \langle \op O \rangle_{\mathcal F_1} + \langle \op O \rangle_{\mathcal F_2}$.
This fact enables us to define the fragment projection matrix from, for
example, atomic projectors, even when, as in most of the cases, the atoms cannot be
considered as pure system moieties.

Instead of Eq.~\eqref{fragexpval}, we could have defined the fragment expectation
value by the equation:
\begin{equation}
\langle \op O \rangle_{\mathcal F} = \trace{ \op W^\mathcal F \op F \op W^\mathcal F \op O} = \trace{ \op F \op O^\mathcal F}\;,
\end{equation}
which we know for a pure fragment would have lead to the same result.
This shows that, even in the case of an operator that is not fragment-block diagonal,
for a pure fragment only the diagonal term contributes to the expectation value,
which is a natural result of the use of Hermitian operators.

\subsection{Population Analysis of Fragments}
\label{sec:FMaP}
Within this framework, traditional population analysis schemes might be extended
to a system's fragments. In Ref.~\cite{mohr-fragments1-2017} we introduced
expressions for the Mulliken ($M$) and L\"owdin ($L$) projectors, which in the
basis representation are:
\begin{equation}
 \mathbf{R}_M^\mathcal F \equiv \mathbf T^\mathcal F \mathbf S^{-1}\;,\quad
 \mathbf{R}_M^\mathcal F \equiv \mathbf S^{-1/2} \mathbf T^\mathcal F \mathbf S^{-1/2}\;,
\end{equation}
where $T^\mathcal F$ is a diagonal matrix which has a value of one for the indices
$\alpha \in \mathcal F$ that are
associated to the fragment $\mathcal F$. Such an association is somehow arbitrary,
in the sense that it is based on simple geometric considerations on the domain
of the basis functions. The value of $\Pi_\mathcal F^{M,L}$ enables one to
assess whether the fragmentation is reliable within the
chosen population scheme.
Also, the matrix $T^\mathcal F$ may be expressed as:
\begin{equation}
 T^\mathcal F = \sum_{a \in \mathcal F} T^a\;,
\end{equation}
where we define the matrices $T_a$ by associating each index $\alpha$ to
one atom $a$ of the system. We retrieve in this way the traditional
Mulliken and L\"owdin atomic projections. The well-known unreliability of
these population methods for atoms may the be ascribed to the fact that,
in general, the atoms cannot be associated to pure fragments: in most of the
cases $\Pi_a$ would be significantly different from zero.

\section{Significance of the Purity Indicator}
\label{sec:SigofPI}
We may give to the purity indicator a chemical significance.
Indeed, given a basis set and a projection method, the orbital population of a fragment can be defined as follows:
\begin{equation}
  q_{\mathcal F} \Pi_\mathcal F =
  \langle \op F ^\mathcal F \rangle_{\mathcal F} - \langle \identity  \rangle_{\mathcal F} = \mathcal B_\mathcal{FF}  - \langle \identity  \rangle_{\mathcal F}\;.
\end{equation}
In the above definitions we have employed the \emph{orbital population} of the fragment
$\mathcal F$, defined as $\langle \identity \rangle_{\mathcal F} = \trace{\mathbf{ K S}^\mathcal{F}}$,
as well as the \emph{fragment bond order}, which is a quantity that in general involves
two fragments:
\begin{equation}
  \mathcal B_\mathcal{FG} =
\Tr\left( \mathbf{ K S}^\mathcal{F} \mathbf{ K S}^\mathcal{G} \right)
  = \langle \op F ^\mathcal G \rangle_{\mathcal F}\;.
\end{equation}
Such a quantity is associated to the overall bonding ability of the two fragments $\mathcal F$ and $\mathcal G$ \emph{with respect to the chosen basis set and population scheme}.
This quantity is similar to the Wiberg index~\cite{wiberg1968application},
and in the case of the Mulliken representation with atomic fragments corresponds to the
Mayer bond order~\cite{mayer1984bond}. In this case, we have defined a
more general fragment bond order, which describes the interaction between
two arbitrary fragments.

The purity condition defined in this way strongly resembles the concept of
chemical valence~\cite{borisova1973molecular, armstrong1973bond},
which measures the ability of an atom to form chemical bonds in its current
environment, but in this case we include off diagonal contributions and scale
by the number of electrons.
Indeed it is enough to notice that, in the Mulliken population scheme,
for a fragment made only of atom $a$ the purity indicator is the
opposite of the atomic valence:
\begin{equation}
    q_a \Pi_a = - V_a \equiv = \trace{ \left(\mathbf{K S T}^a\right)^2 - \mathbf{K S T}^a}.
\end{equation}

Following this interpretation we can rephrase the purity condition with a
chemical meaning: a fragment is pure if it has a ``zero-valence'' condition -
i.e.\ the value of the fragment bond order with itself equals the fragment orbital population.
Despite its physico-chemical interpretation, such a zero-valence condition is a property of the computational setup
and of the projection method, and it is not a chemical property \emph{per se}; however, when the basis set and the projection scheme are suitably chosen,
it enables the splitting of the system's observables into fragments.

As mentioned, the purity indicator has a nonlinear behaviour with respect to the
combination of fragments. It is easy to verify that, for two fragments
$\mathcal F$ and $\mathcal G$ we can expand the purity indicator in terms of
the fragment bond order as follows:
\begin{equation}
\label{eq:environment-pi}
 q_{\mathcal F + \mathcal G} \Pi_{\mathcal F + \mathcal G} =
q_{\mathcal F} \Pi_{\mathcal F} + q_{\mathcal G} \Pi_{\mathcal G} +
\mathcal B_\mathcal{FG} + \mathcal B_\mathcal{GF},
  \;
\end{equation}
such a result will turn out to be useful in the forthcoming section.

\section{Fragmentation and Subsystems}
\label{sec:fragandsub}

We have seen that the fragmentation operators are useful to identify
pseudo-observables that can be associated to a system's moieties.
This is clearly helpful in characterizing a system, providing information
on the impact a given fragmentation will have on the reliability of a
fragment's expectation value.
However, for certain observables, it would be nonetheless interesting
to rely on moieties which are defined beforehand, and analyse their
mutual interaction in order to characterize the system's building blocks
from an electronic point of view.

Let us consider an example scenario where a given target set of atoms
$\mathcal T$ is chosen a priori, and the goal is to compute its properties using
only a subset of the full system. Associated with that target fragment is a
purity indicator $\Pi_\mathcal T$ with a absolute value that may be
higher than some desired threshold $\epsilon$.
We have seen that this implies that the density matrix is not assumed block-diagonal in the fragmentation provided by
$\mathcal T$. Let us define the embedded purity indicator
$ \Pi_{\mathcal T : \mathcal E}$ as the purity indicator of the joint
$T$ and $E$ system, but without considering the contribution associated to the
environment alone. 

\begin{figure}[htp]
 \includegraphics[width=3.2in]{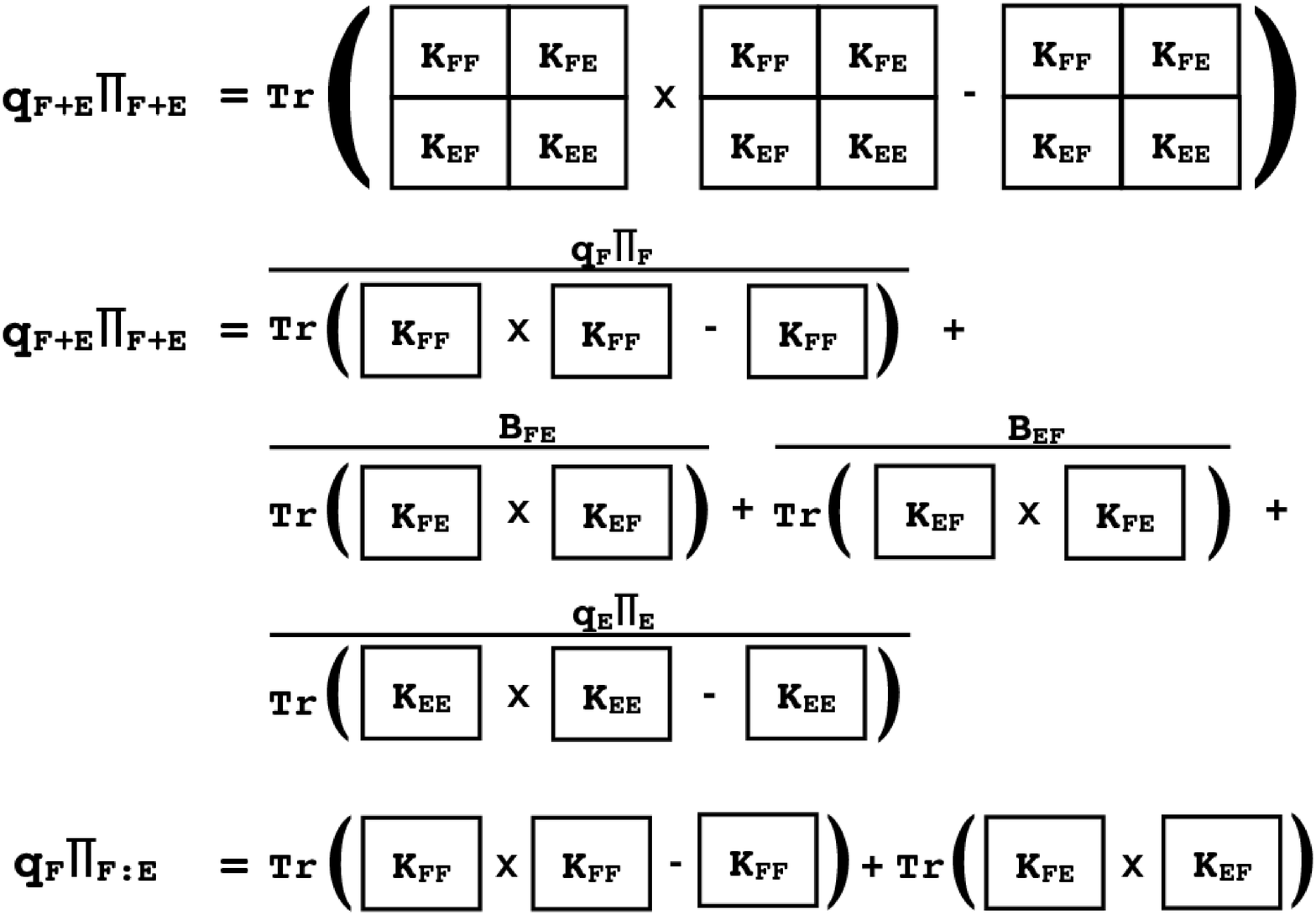}
 \caption{Summary of an expansion of the purity indicator of two fragments
 $\mathcal F$ and $\mathcal E$ in terms of the fragment bond order. For
 simplicity, we assume the Mulliken (or L\"owdin) population scheme with a overlap matrix $S$ that is unitary, but in the general
 case the above diagram is the same with the matrix $K$ replaced by $KS$.}
 \label{fig:puritydiagrams}
\end{figure}

We can define the embedded purity indicator as follows:
\begin{equation}
\label{eq:embedded-pi}
 q_{\mathcal T} \Pi_{\mathcal T : \mathcal E} \equiv
q_{\mathcal T} \Pi_{\mathcal T} + \mathcal B_\mathcal{TE}
  \;.
\end{equation}

To clarify the interpretation of this quantity, in Fig.~\ref{fig:puritydiagrams} we provide a matricial representation of this block
view of the purity indicator.

We note that the correction term
$ \mathcal B_\mathcal{TE} $
represents the ``strength'' of the ``electronic'' interaction between the two
fragments. Crucially, the value of $\Pi_{\mathcal E}$ is not included, with
the trace only running along the $\mathcal F \mathcal F$ block.
Thus, a good environment need not satisfy the purity condition itself.
A suitable embedding environment is one such that the sum of the fragment
bond order values of all fragments excluded from the environment is below some
cutoff. In general, this environment might also be split into a
number of different fragments.

By defining a fragmentation procedure and embedding scheme, we see that a graph
like view of a system emerges. In this representation, fragments are nodes,
and edges are drawn between fragments in the same embedding environment. This
representation can be efficiently computed using the results of a calculation of
the full system. Through judicious choices of a fragmentation and embedding
cutoffs, a coarse grained view of large complex systems can be achieved.

\subsection{Automatic Fragmentation}
\label{sec:AF}
In Sec.~\ref{sec:FaIO}, we established the purity indicator
$\Pi_{\mathcal F}$ as a means of quantifying the choice of a given fragment
$\mathcal F$. With a figure of merit established, we now consider how to
partition a system such that each fragment fulfills that criteria. Determining
the best fragmentation of a system is ill defined as presented so far, as
several different fragmentations of the same system can fulfill the purity
condition. Additional constraints must be introduced, such as locality in
space, similarity to other fragmentation schemes, uniformity in fragment size,
or maximizing the total number of fragments.

For the purposes of this paper, we will consider a simple greedy, spatially
motivated algorithm for fragmenting the system. We begin by treating each
atom as its own separate fragment. Then, we select the fragment with the
lowest purity value to be merged. The fragment bond order between this
fragment and its neighbors within a 10 Bohr radius are computed, and we merge
it with the fragment with the largest bond order. This process is repeated
until all fragments satisfy the purity condition
$\Pi_{\mathcal F} > \epsilon$. While this fragmentation is not
guaranteed to maximize the number of fragments, it is efficient to compute,
and the spatially local fragments will help with subsequent analysis.

\section{Reliability of the Approach}
We will now demonstrate the previously presented tools on a number of example systems.
We consider four example systems: a cambrin protein
(\textsc{1CRN})~\cite{teeter1984water}, a Laccase enzyme from Trametes versicolor
(\textsc{Laccase})~\cite{dellafiora2017degradation}, a cluster of pentacene molecules (\textsc{Pentacene}),
and an RNA molecule binding magnesium (based on PDB 1I7J~\cite{adamiak20011})
in solution (\textsc{MG}) (see Fig.~\ref{qmmmimages}). Details of the DFT calculations performed are presented in Sec.~\ref{sec:DFTdet}
of the Appendix.
These systems each represent different challenges for complexity reduction.
\textsc{1CRN} is a well studied model, and coarse graining of the
system might be achieved by simply decomposing the fragments based on the
amino acid sequence. \textsc{Laccase}, on the other hand, has four copper
atoms in it, making it not possible to decompose it purely using the amino
acids. For the \textsc{Pentacene} system, the fragmentation guidance is somehow obvious,
and it is interesting to decompose the observables into bulk-like and surface fragments.
For the \textsc{MG} system, while partitioning of water molecules is an
obvious start, whether the RNA molecule can be partitioned remains uncertain,
as is determining a suitable fragmentation for the  magnesium ions. For all of
these systems, even once a decomposition has been established, the choice of
an embedding environment for each target fragment remains challenging.
\begin{figure*}[!bt]
 \centering
 \begin{tabular}{cc}
 \subfloat[][\textsc{1CRN}]{\includegraphics[width=3.2in]{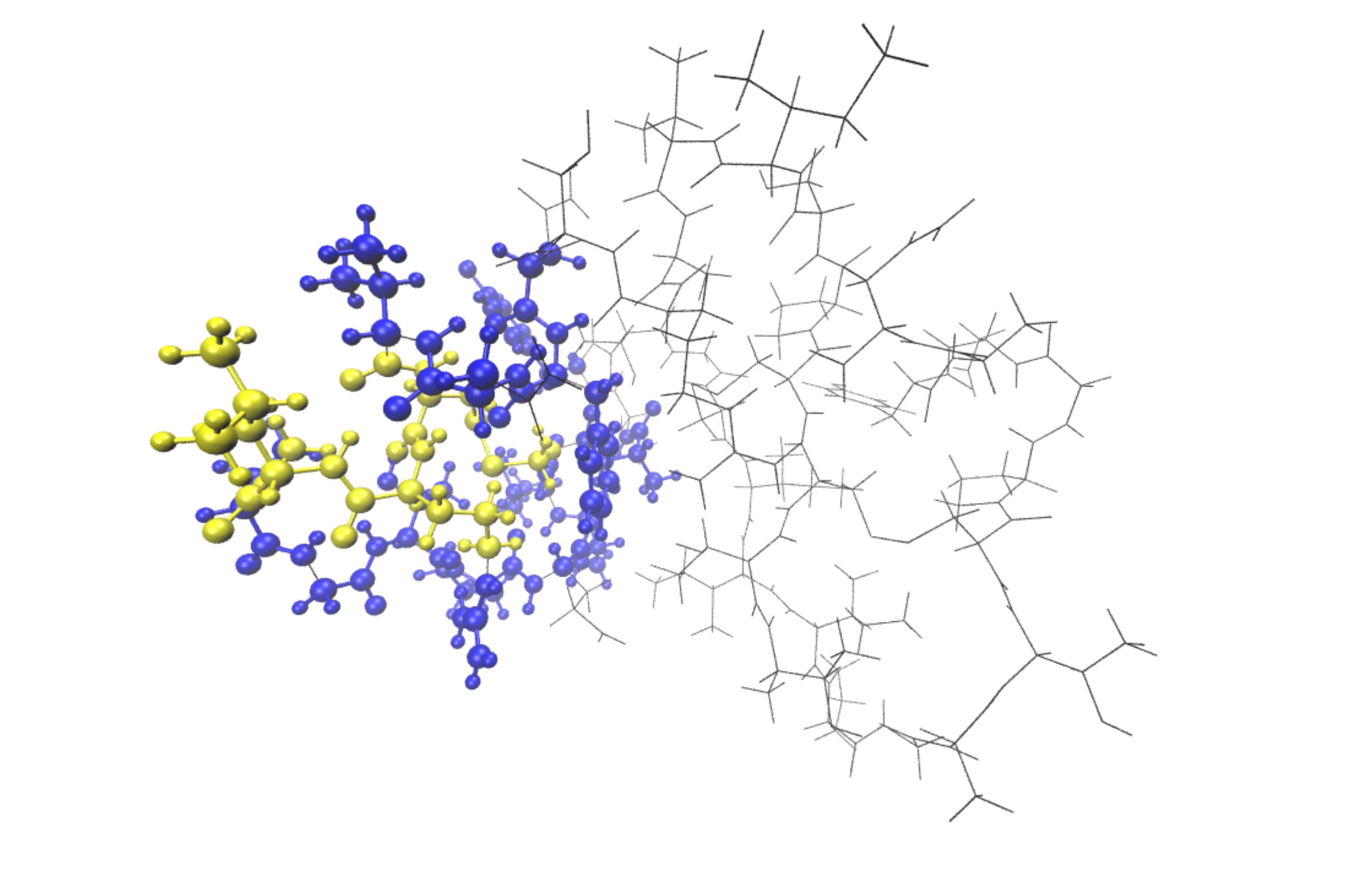}\label{fig:1crn-qmmm}} &
 \subfloat[][\textsc{Laccase}]{\includegraphics[width=3.2in]{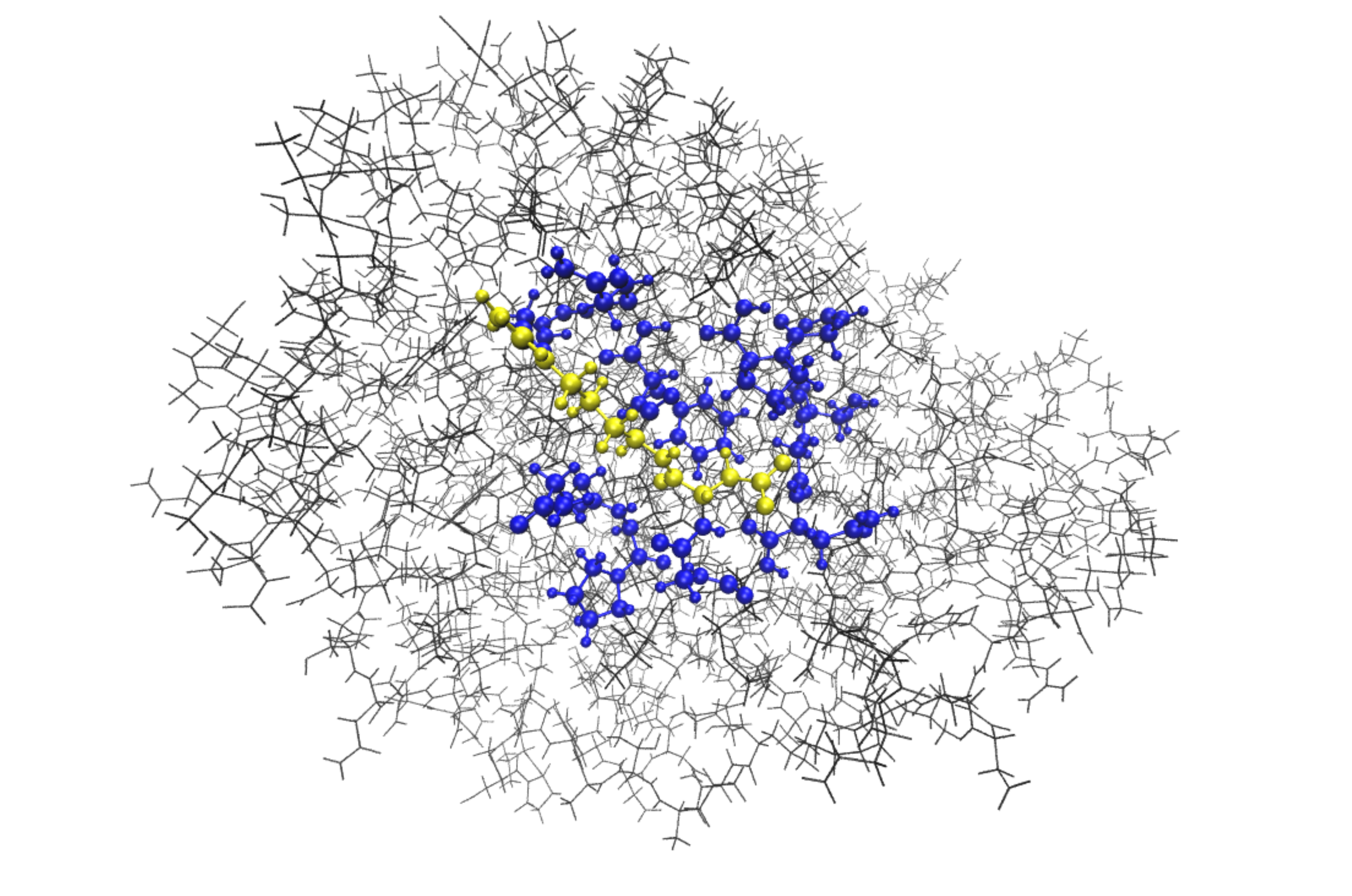}\label{fig:Laccase-qmmm}} \\
 \subfloat[][\textsc{MG}]{\includegraphics[width=3.2in]{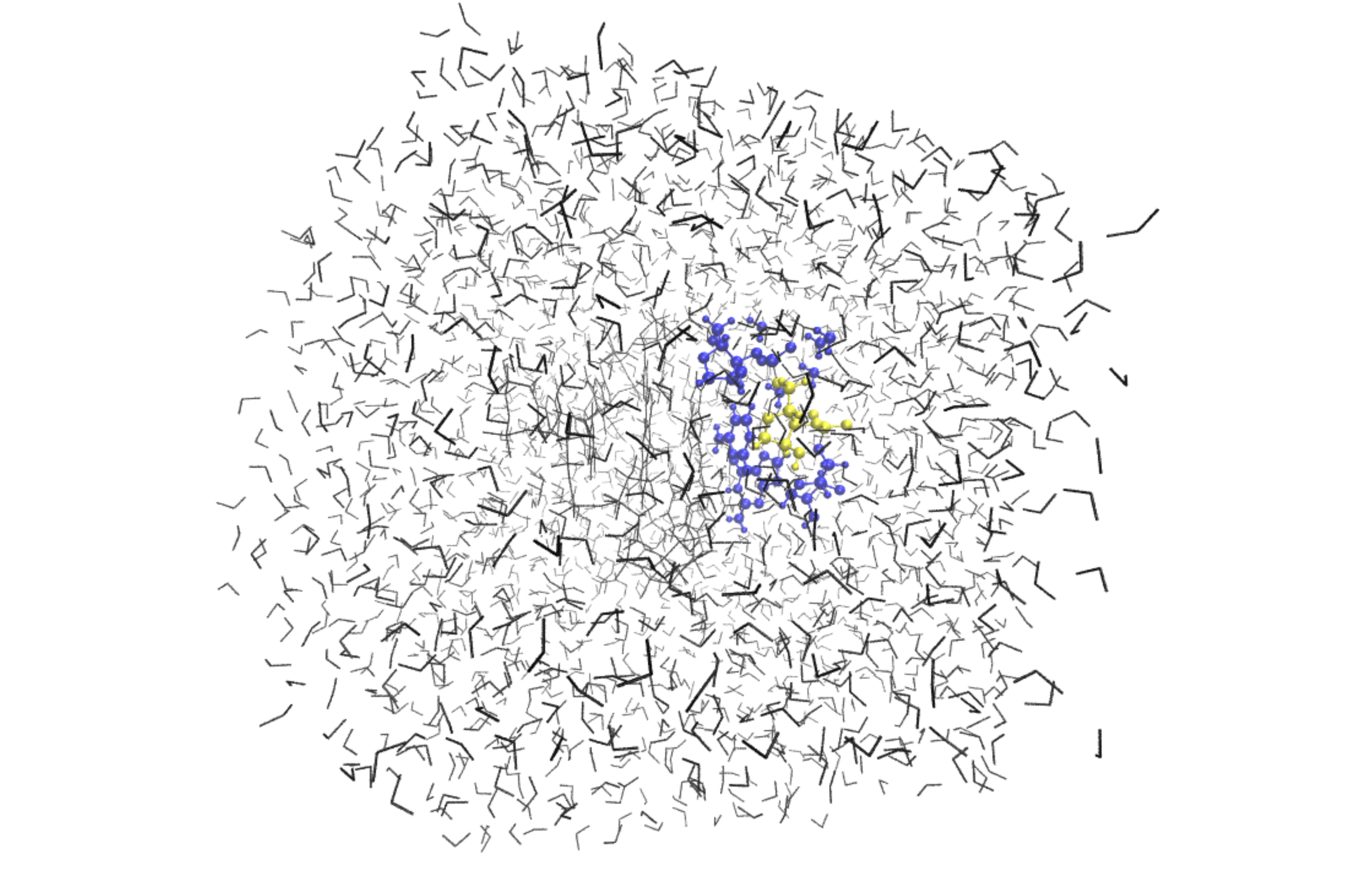}\label{fig:MG-qmmm}} &
 \subfloat[][\textsc{Pentacene}]{\includegraphics[width=3.2in]{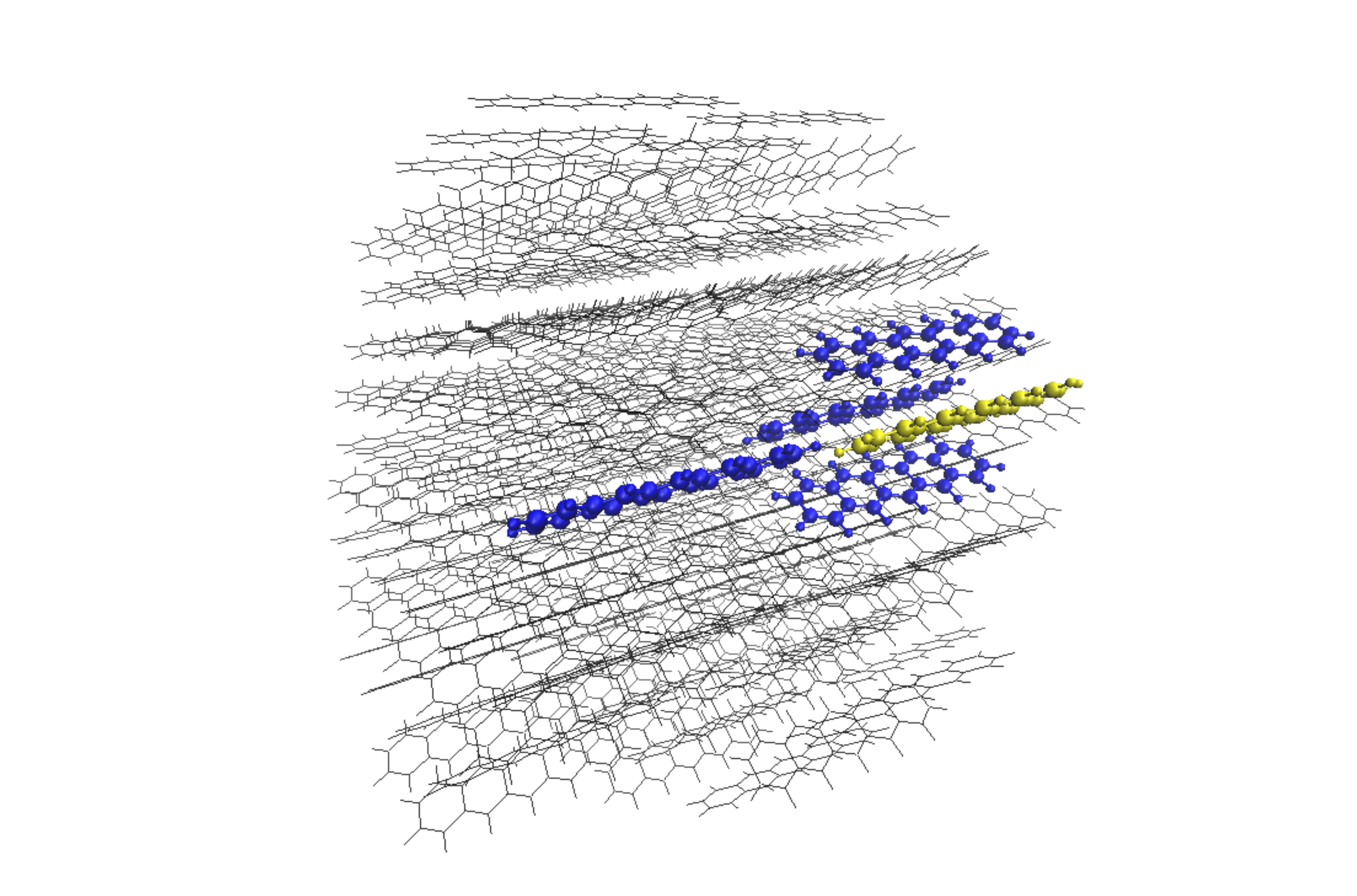}\label{fig:Pentacene-qmmm}}
 \end{tabular}
 \caption{Embeddings of target fragments in the four sample systems. The
 target regions are in yellow, and the embedding environment (using a bond
 order cutoff of $0.01$) are in blue. Atoms in black are those which belong to
 the full system but are excluded from the subsystem calculations.\label{qmmmimages}}
\end{figure*}

The tools established in the preceding sections require no a priori information
about the system to be applied, and can generate an unbiased coarse graining of
each type of system. In this section, we will systematically fragment
and compute embedding environments for these example systems, and evaluate these
reduced models with a number of different metrics.

\subsection{Choice of Purity Indicator Cutoff}
\begin{figure}[htp]
\includegraphics[width=3.2in]{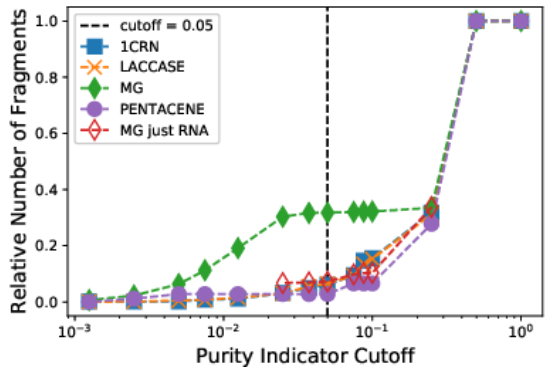}
\caption{The relative number of fragments in each system compared to an atomic
fragmentation at various purity indicator cutoff (absolute) values.\label{fig:numfrags}}
\end{figure}

We begin by exploring the choice of purity indicator cutoff's effect on the
number of fragments in a given system. For each system, the auto fragmentation
procedure described in Sec.~\ref{sec:AF} is applied. The number of fragments
for each system at various cutoffs are plotted in Fig.~\ref{fig:numfrags}. We
have also analyzed the number of fragments of just the RNA molecule in
the \textsc{MG} system.

One point of interest in the  data of Fig.~\ref{fig:numfrags} is that the two proteins
(\textsc{1CRN}, \textsc{Laccase}) follow an extremely similar
trend when comparing the relative number of fragments at a given cutoff value.
This suggests that the average size of a fragment is similar when systems
are composed of similar building blocks. This is in contrast to the
\textsc{Pentacene} system which has similarly sized fragments, but different
building blocks with different behavior. As a further point of contrast, the
\textsc{MG} system has relatively more fragments at high cutoff values than
all of the other systems.  However, when separately examining the RNA molecule fragments in
\textsc{MG}, we see a different picture, with a stricter cutoff leading to fewer
fragments.

There appears to exist a region between $\Pi=-0.01$ and $-0.05$ where the
number of
fragments is relatively stable, and a coarse grained view of the system is
possible. We note that this regime matches a similar finding as a study
using localized orbitals to partition domains for the purpose of accelerating
exact exchange calculations~\cite{dawson2015performance}. In that study, the
cutoff was defined in terms of the norm of truncated localized orbital,
which corresponds closely with the purity indicator. 

For a given cutoff value,
there exists some freedom based on how coarse grained a view of a system is
desired. For example, with the \textsc{MG} system, a cutoff in this range
may be too fine grained a view of the system, leading to a tighter cutoff
value for fragmenting the solution. This is due to the large number of water molecules in
the system. Water molecules have very low (absolute) purity values, meaning
that a fragmentation of a solution is stable even as the cutoff value is
tightened.

For the remainder of this paper, we will use a cutoff of $\Pi=-0.05$ for
automatically fragmenting systems. Further analysis will be performed on this
choice of cutoff in Sec.~\ref{sec:etcgv}.

\subsection{Choice of Fragment Bond Order Cutoff and Reliability of Fragment Observables}
We now consider the appropriate threshold for defining an embedding environment.
As a figure of merit, we focus our attention to the 
electrostatic dipole as the chosen fragment observable (the operator $\hat O$ of the equations in 
Sec.~\ref{sec:FaIO}).
We have chosen the dipole since a faithful representation of such observable 
would demonstrate the reliability of the electronic density as well as a good approximation 
for one of the main quantities needed for the long-range potential that a fragment would generate.

To do this, we begin by fragmenting each of the example systems, with a cutoff
of $\Pi=-0.05$. Next, for each system we select the fragment with the largest
dipole value (to increase the signal to noise of subsequent calculations) and
define it as the target fragment for embedding. We then define embedding
environments based on various threshold values.
We also compared this approach with
an environment computed by the nearest neighbor distance between fragments.

Calculations were then performed from scratch on the target and embedding
environment, and observables were recomputed.
The dipoles of each target fragment
in the various embedding environments were computed from the atomic dipoles
according to the equations in our previous
publication~\cite{mohr-fragments1-2017}.
We emphasize that no external potential from
outside the embedding environment was included except through the net charge
which was rounded to the nearest electron. 

\begin{figure}[htp]
\includegraphics[width=3.2in]{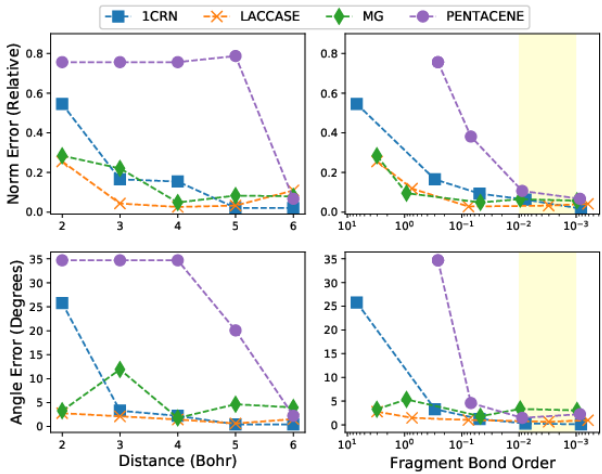}
\caption{Error in the dipole in various embedding environments. The relative
error in norm is defined as $\frac{||d' - d||_2}{||d||_2}$ where $d$ is the
dipole computed from the full calculation, and $d'$ the dipole computed
from the subsystem calculation. The angle error is the angle between the
dipole computed from the full and subsystem calculations. The region between
a fragment bond order cutoff
of $0.01$ and $0.001$ has been highlighted to emphasize the converged
observables.\label{fig:dipole-error}}
\end{figure}

Images of the various target
regions inside an embedding region with a bond order cutoff of $0.01$ are shown
in Fig.~\ref{qmmmimages}.
Errors in the dipole values
are plotted in Fig.~\ref{fig:dipole-error}. 

We note that it is remarkable that these calculations are
accurate at all given that in many places we have cut covalent bonds without
using a capping procedure. Calculations performed on these systems also
smoothly converged to the ground state. The lack of a need for capping can be
attributed to the $\Pi=-0.05$ purity value of the fragments, which already
limits the amount of charge being leaked. In a sense, the low (absolute) purity
value of the embedding fragments enable them to act as a general type of cap on
the target fragment. By adding the embedding environment, it is possible to
significantly reduce the errors, until improvement stagnates in general with a
bond order cutoff of between $0.01$ and $0.001$. 
We also note that while a conservative distance criteria can define a suitable
embedding environment, the converged distance
value is significantly affected
by the specific system geometry. Using the auto fragmentation procedure and
bond order tools together, one can automatically define an embedding of all
system fragments which accurately reproduces desired observables.
\begin{figure}[htp]
\includegraphics[width=3.2in]{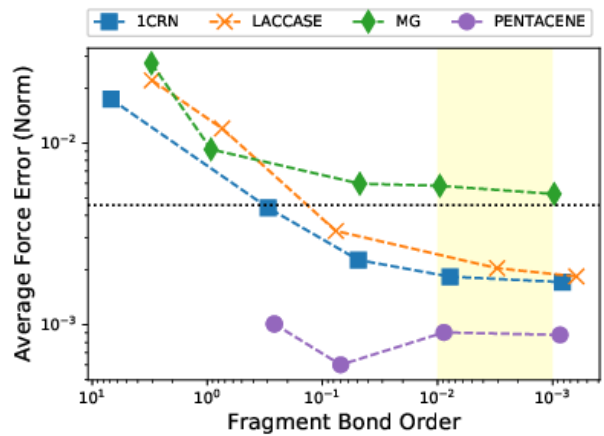}
\caption{Average error in the atomic forces (a.u.) in various embedding environments. The dotted horizontal line is the estimate of the noise in the force from the calculation of the full \textsc{1CRN} system.\label{fig:force-error}}
\end{figure}

We also consider the error in the atomic forces. We know that, as 
the atoms cannot be associated to pure fragments,
an embedding environment is needed to guarantee that atomic 
forces can become reliable. We have computed 
the average error in the forces inside the target fragment and 
plotted those values in Fig.~\ref{fig:force-error}.
Here we see a similar convergence trend, with the exception being 
the  \textsc{Pentacene} system, which has very low forces on any 
given fragment. To put these errors into context, we define
an estimate of noise in the forces as the standard deviation of the forces with mean zero.
The  error in the forces presented becomes of the same order of magnitude as the noise in the
forces from calculations of the full system, and are similar to the errors in forces that come 
from using the linear scaling version of BigDFT~\cite{mohr-accurate-2015}. The
stagnation in force error reduction with an increasing environment size is
further evidence that we have captured the essential fragment environment for
all systems using a bond order cutoff between $0.01$ and $0.001$.

For a fragment identified with a $|\Pi| = 0.05$, and using a bond order cutoff value
of $0.01$, the number of atoms in the four system target regions are
228, 191, 97, and 180 for \textsc{1CRN}, \textsc{Laccase},
\textsc{MG}, and \textsc{Pentacene} respectively. The
embedding environments contain 172, 159, 78, and 144 atoms. If these regions
were used for production QM/MM calculations, they would represent larger QM
regions than are usually treated, though recent studies have favored bigger
QM regions~\cite{cole2016applications}. A reduction of the QM region size is
possible when the MM region realistically mimics the external region, such
as through the inclusion of capping atoms or the use of well tuned MM
potentials.

The procedure presented here requires no direct user interaction, and is instead a general workflow for
studying any kind of system. This generality is further shown in the
supplementary information, as the calculations on each system can be
performed with the same script by only changing the input geometry file. The
generality of this scheme makes it a promising approach for
high-throughput calculations aimed at complexity reduction.

\section{Evaluating The Coarse Grained View of the System}
\label{sec:etcgv}
We now continue our analysis on these systems by performing a more information
centric analysis of the system fragments. We will begin by studying the 
transferability of fragments, and comparing them to the amino acids of proteins.
We will then generate graph like views of each system, and see how choices of
fragment purity and embedding environment affect various graph metrics. 

\subsection{Fragmentation Comparison of Proteins}
\begin{figure}[htp]
\includegraphics[width=3.2in]{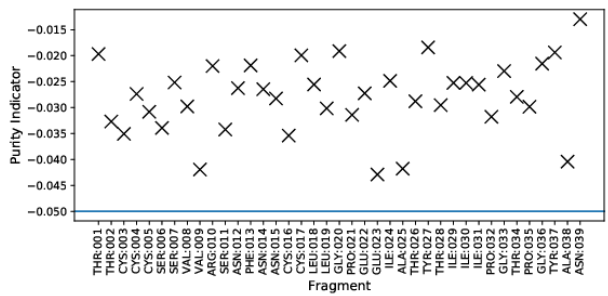}
\caption{Purity indicator values for the \textsc{1CRN} system when fragmented
by amino acid using the FU program.\label{fig:1crnfrags}}
\end{figure}

For \textsc{1CRN}, a different natural fragmentation might be to use the
amino acid sequence of the protein instead of the auto fragmentation procedure.
We have computed the purity values of those
fragments as generated by the FU program~\cite{fedorov2017modeling}, and
plotted them in Fig.~\ref{fig:1crnfrags}. We see that by our purity criteria
of $\Pi > -0.05$, the amino acids are a reasonable system fragmentation, 
which is not surprising
for this kind of model system. Nonetheless, the auto fragmentation procedure
requires no a priori fragment information, making it applicable to a wider
class of systems. When additional fragmentation guidance is available, the two
approaches can be combined if a coarser grained view of the system is desired.
For example, with the \textsc{1CRN} system, tightening the threshold from
$\Pi=-0.05$ to $-0.025$ to $-0.01$ reduces the number of fragments from 39 to 18
to 5. When starting from the amino acids for the \textsc{Laccase} system, the
drop is from 452 to 215 to 77 fragments.

\begin{figure}[htp]
\includegraphics[width=3.2in]{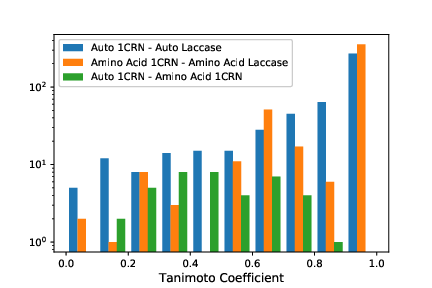}
\caption{Histogram of Tanimoto coefficients when comparing the fragments of
\textsc{1CRN} with \textsc{Laccase}. Note that Tanimoto coefficient values
range from $0$ to $1$, with values closer to $1$ being more similar.\label{fig:tanimoto}}
\end{figure}

This result further demonstrates that some arbitrariness exists in the choice
of system fragmentation. This might hint that there exist a broader set of
descriptors for describing biological systems than just the amino acid
sequence. Using the open babel code~\cite{o2011open, o2008pybel}, we can for each
fragment compute a molecular fingerprint, and then compute a similarity
score between each pair
of fingerprints. For this study, we will use the FP2 fingerprint, which
creates a binary string representation of a fragment based on short linear
and ring molecule substructures, and evaluate the similarity between those
strings using the Tanimoto coefficient (see Willet~\cite{willett2006similarity}
for an overview of this approach). In Fig.~\ref{fig:tanimoto}, we demonstrate
this approach by first comparing the fragments of \textsc{1CRN} using the
auto fragmentation tool and the amino acid sequence. We see that these
fragments are indeed significantly different, despite the fact that both the
amino acid partitioning and the auto fragmentation procedure result in the
same number of fragments (39). Next, we investigate the transferability of
fragments by comparing the fragments of \textsc{1CRN} with \textsc{Laccase}.
For \textsc{Laccase}, it is difficult to determine an appropriate fragmentation for the
copper atoms without a tool like the auto fragmentation procedure, but for this
comparison we have by hand merged all the copper atoms with
their neighboring cystine amino acids. When comparing the amino acids of
\textsc{1CRN} and \textsc{Laccase}, we unsurprisingly identify many similar
fragments. However, we also compare the fragments of \textsc{1CRN} generated
with the auto fragment tool, and find that there are also many similar
fragments shared between the two systems. Thus, while the auto fragmentation
tool identifies new kinds of fragments, these fragments remain transferable,
making them a promising source of new descriptors that can more adequately be put in relation 
with QM calculations in a given computational setup.

\subsection{Graph Metrics on General Systems}
We finish this demonstration by turning to the generation of graph like views
of a system. For each of the example systems, we once again perform auto
fragmentation with a $\Pi=-0.05$ cutoff and use this fragmentation to define the
graph's nodes. Then, for each fragment, we compute its embedding environment
at various thresholds, and use that environment to define the edges of the
graph.

\begin{figure}[htp]
\includegraphics[width=3.2in]{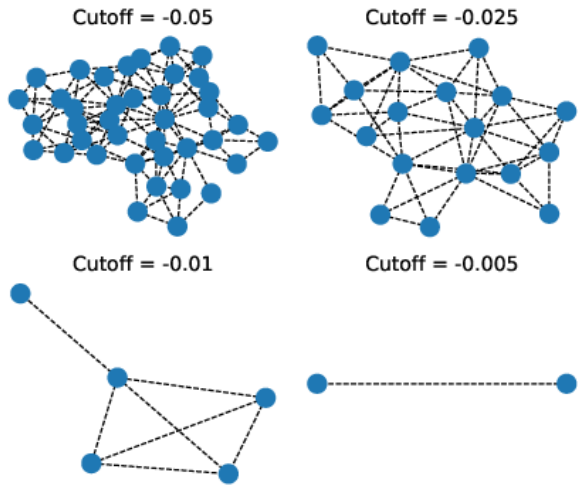}
\caption{Coarse graining of the \textsc{1CRN} graph structure starting with the
amino acids fragments. The diagrams here are generated using the Kamada-Kawai
algorithm~\cite{kamada1989algorithm} for visual clarity, so node locations
are not related to atomic positions in space.\label{fig:pidrop}}
\end{figure}

We may examine the graph characteristics of a system with a change in purity indicator
cutoff while keeping the bond order cutoff fixed. By increasing the purity
cutoff closer and closer to zero, we can generate low resolution views of
a system's connectivity. This process is demonstrated in Fig. \ref{fig:pidrop}.
In this example, we begin with the connectivity of the \textsc{1CRN} system
with the fragments defined by the amino acids and the connectivity with a
$0.01$ bond order cutoff. As we push the cutoff closer to zero, the shape of
the graph changes significantly, resulting in a simpler and simpler picture
of the system.

From this representation, we compute some sample graph metrics: the average
shortest path length and the average clustering coefficient~\cite{saramaki2007generalizations}.
These metrics have been applied to proteins in the past, as reviewed by Estrada~\cite{estrada2012structure}. In most previous studies, however, the focus was on long
range van der Waals interactions. Other authors have used interaction energies
such as those derived from forcefield calculations~\cite{vijayabaskar2010interaction}. 
Sladek and coworkers recently utilized pair interaction energies~\cite{fedorov2007pair} 
to define network edges, and showed how the properties analyzed using
an energy based model differ from standard distance based analysis~\cite{sladek2018protein}.

\begin{figure}[htp]
\includegraphics[width=3.2in]{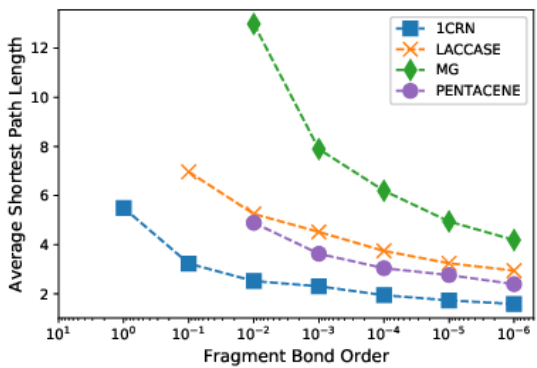}
\caption{Average shortest path length for each system at various cutoff
values. The path length of a disconnected system is $\infty$, so those values are not shown here.\label{fig:aspl}}
\end{figure}

Values of the average shortest path length metric are reported in
Fig.~\ref{fig:aspl}. Note that in the analysis presented here, the average shortest
path length is defined in  terms of the number of edges traversed, without consideration
to physical inter-fragment distances.
From this figure, we find additional supporting evidence
for a bond order cutoff of $0.01$ for the embedding environment.
When a smaller value is used, the graphs of these systems are no longer
fully connected. Even with a bond order cutoff value of $0.1$, the \textsc{MG}
system is disconnected, reflecting how pure the water molecule fragments are.
The two proteins are connected as soon as any bond order is considered, but the
average shortest path length quickly decreases with a decrease in
fragment bond order cutoff, leading to a very different description of the
system.

\begin{figure}[htp]
\includegraphics[width=3.2in]{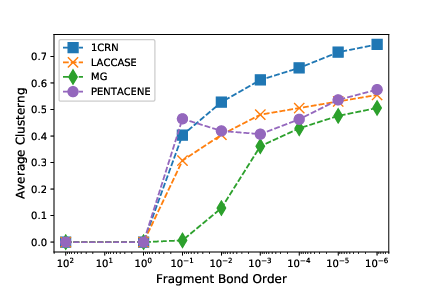}
\caption{Average clustering coefficients for each system at various cutoff
values.\label{fig:acc}}
\end{figure}

The average clustering coefficients are plotted in Fig.~\ref{fig:acc}.
The low average clustering value for the \textsc{MG} system reflects the
general lack of structure of the water molecules, while the other systems
are significantly more connected. For both of
these metrics, we find inflection points with a bond order cutoff of around
$0.01$ or $0.001$, after which the measures increases/decrease linearly with
the fragment bond order. However, the slope of the linear region depends on the
system. Thus, while networks might be generated with large distance cutoffs to
incorporate long range interactions, the fundamental structure of graphs can be
understood by looking at the short ranged covalent interactions using the
fragment bond order tool.

\begin{figure}[htp]
\includegraphics[width=3.2in]{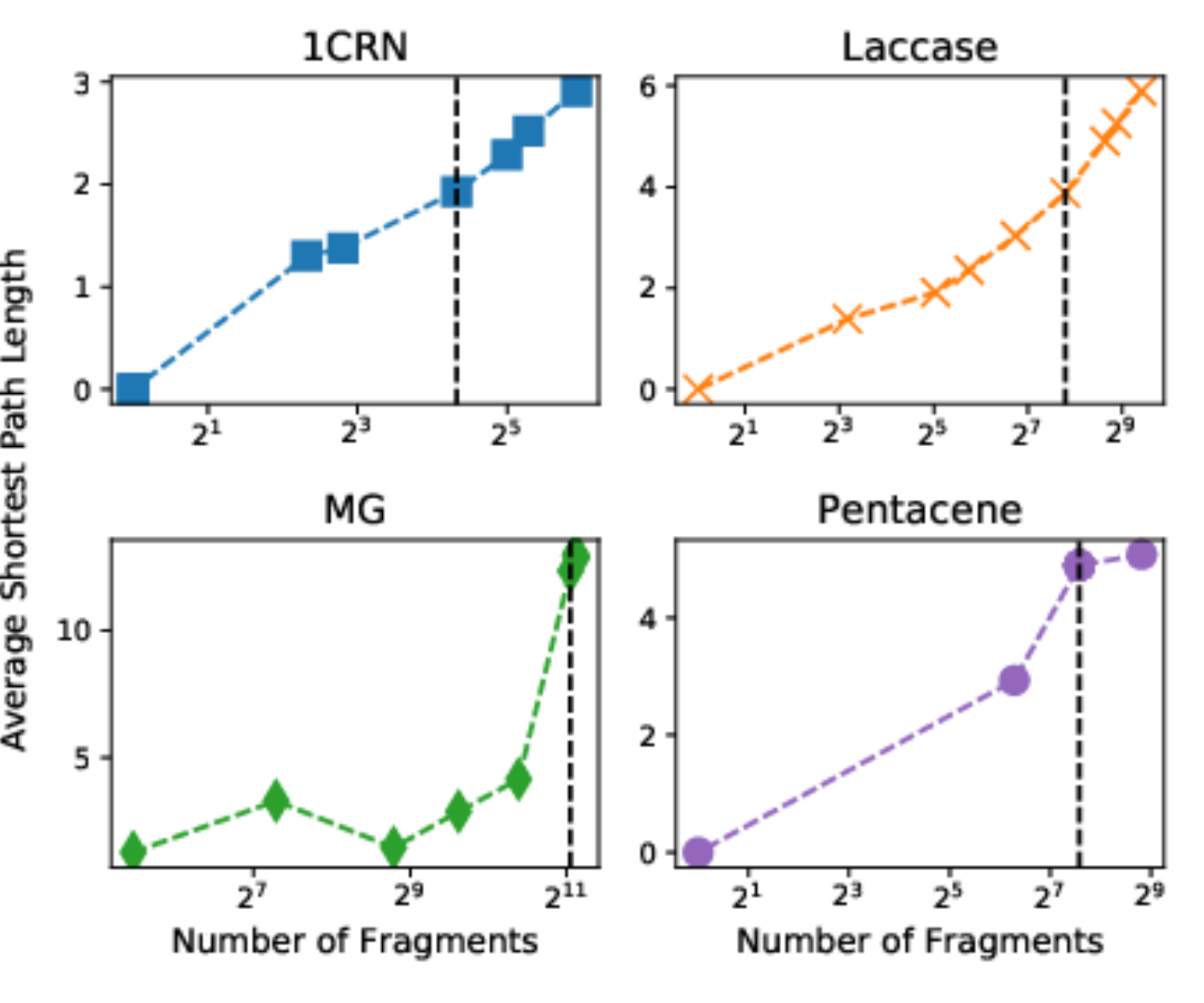}
\caption{The average shortest path length vs. the number of fragments for
each system at various purity value levels. The dotted vertical line represents
the number of nodes when a purity value of $\Pi=-0.025$
is used.\label{fig:asppi}}
\end{figure}

Next, we apply this approach to all four example systems using the fragments
defined by the auto fragmentation procedure. For each of these systems, we
compute the average shortest path length at various purity cutoffs, and plot
it against the log of the number of fragments in Fig. \ref{fig:asppi}.
For the \textsc{MG} system,
certain values of the network were disconnected, so an average was taken
over each subgraph. For a network with small world
characteristics, the average shortest path length should grow logarithmically
with the number of nodes~\cite{barrat2000properties}.
Intriguingly, we do see such growth for the two
protein molecules, though with an inflection point around a purity indicator
cutoff of $\Pi=-0.025$. For the other two systems, there also appears to be two
distinct patterns centered at a purity value of $\Pi=-0.025$. This suggests
that there is a cross over point at which a view of the local structure is lost
and the global structure dominates the description of the molecular system. In
the complexity reduction framework proposed here, this information can be
extracted using the BigDFT code, enabling insight into system properties at the
desired level of detail.

\section{Fragmentation vs. Embedding}
In the framework introduced thus far, users have two degrees of freedom to
consider for building a coarse grained model of a system. The first is the
coarseness of the fragments, as determined by the purity indicator. The second
is the choice of embedding environment as defined by the fragment bond  order.
The balance between these two variables depends on the choice of observable one
wishes to compute, as we will demonstrate through the following case study: 
computing the density of states (DoS) of the \textsc{Laccase} system.

\begin{figure}[htp]
\includegraphics[width=3.2in]{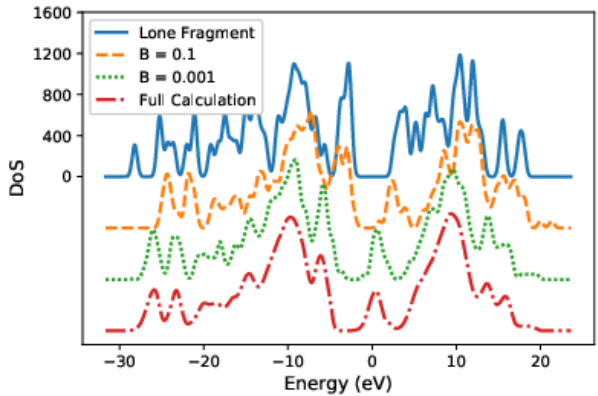}
\caption{The projected density of states of the target region of the
\textsc{Laccase} protein computed with different embedding environments. 
Plot was obtained using a smearing parameter of 0.3 eV and each line has been shifted for visual clarity.}
\label{fig:pdos}
\end{figure}

First, we consider the problem of computing the density of states of a system 
projected on to a given fragment~(PDoS). 
In the framework of fragmentation, the projected density of states can be computed by the 
formula
\begin{equation}
 \rho_{\mathcal F}(\omega) = \sum_i \trace{\op W^\mathcal F \ket{\psi_i}\bra{\psi_i}} \delta( \epsilon_i - \omega)\;,
\end{equation}
where $\ket{\psi_i}$ are the Kohn-Sham orbitals of energy $\epsilon_i$.
It is indeed possible to compute the
PDoS for any arbitrary fragment. However, we know that the DoS of a given fragment
will be influenced by its environment, with the degree of influence determined by the
purity indicator of that fragment. When a fragment is not pure, if we recompute
that fragment in isolation, we can't expect to reliably reproduce the 
PDoS embedded in the full system. However, by including more and more environment in a buffer
region using the fragment bond order as a guide, we can eventually reach a converged
result, as shown in Fig.~\ref{fig:pdos}. Thus, the bond order tool allows us to 
make up for the lack of purity of a given fragment by performing embedding
calculations.

\begin{figure}[htp]
\includegraphics[width=3.2in]{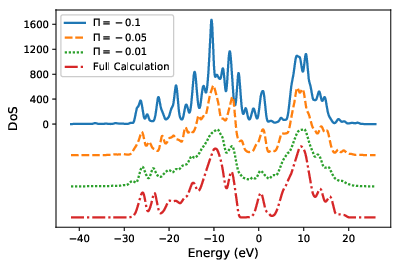}
\caption{The density of states of the full
\textsc{Laccase} protein computed as a composite
from subsystems computed at different purity values.
Plot was obtained using a smearing parameter of 0.3 eV and each line has been shifted for visual clarity.}
\label{fig:dos}
\end{figure}

Now we turn to computing the full DoS of the entire systems. Here we might be
tempted to use the same approach (i.e. for each fragment, we compute its
embedding environment, perform an embedded calculation, compute the projected
DoS, and finally sum up the values). However, an alternative approach would be
to simply use the auto fragmentation procedure with a stricter cutoff, and use no
embedding environment, as shown in Fig.~\ref{fig:dos}. The benefit of an embedding
environment is that the target region has an effective purity value that is
closer to zero than by itself. However, by building the embedding environment,
we also improve the purity value of each of the embedding fragments, resulting
in a total purity value for a joint target-embedding fragment that is much
closer to zero than the lone target. Thus, a non-buffer approach can reduce
the amount of repeated work from overlapping environments, with the trade-off
depending on the scaling of the computational cost with the system size.

The difference between computing the DoS and the PDoS is that in one case we
care about a \emph{system level observable} and in the other a 
\emph{fragment level pseudo-observable}. For each calculation quantity, we should
consider the level of detail it is computed at. The fragment dipole and the PDoS are fragment level quantities, whereas the DoS or the
total energy are system level quantities. The purity indicator gives us
a measure that informs whether such quantities can be computed independent of
the environment. In cases where it is not possible, the bond order tool can
define a suitable embedding such that one reproduces the desired value.

\section{Discussion}
In this work, we have presented a complexity reduction scheme which takes
large, heterogeneous systems, and uses the results of linear scaling
DFT calculations to generate coarse grained models. We have
demonstrated how this approach can be applied without bias to different
classes of systems with no a priori user information. Furthermore, by
applying this method to generate model systems of a target in an environment,
and using these models to accurately compute fragment observables, we have
shown that this approach provides chemically meaningful descriptions
of system interactions.

Fragments form the basis for many low order scaling methods for computing
large systems. Those methods compute the properties of a system from the ground
up, using either predefined fragments (for example, using the amino acid
structure of proteins~\cite{wang2013electrostatically, he2014fragment}),
fragments refined using distances between fragment
elements~\cite{ganesh2006molecular, hua2010efficient, hua2013generalized},
cheminformatics~\cite{steinmann2012fragit}, or bonding information in
combination with chemically motivated rules~\cite{deev2005approximate,
le2012combined, collins2012systematic, collins2014combined} to
define the partitioning. A recent review by Collins and
Bettens~\cite{collins2015energy} describes many of these types of methods.
The approach presented here differs in that it is instead works from the top
down, using the
results of linear scaling calculations to determine system fragments. This work
is thus more focused on post-processing systems for chemical understanding than
on fast calculations. Nonetheless, our approach can serve as a complement for
such methods by defining initial partitioning and embedding systems,
which then can be treated at a higher level or theory, have their geometry
optimized, or used to perform molecular dynamics with such fragment methods.

The interaction between fragments also has been a topic of many studies,
in particular when trying to determine intermolecular forces for studying
reactions~\cite{szalewicz2012symmetry, fedorov2007pair, parrish2014chemical}.
This work goes back to the pioneering
development of the theory of atoms in molecules~\cite{bader1991quantum}, as
the critical points in the electron density can be used to define which atoms
interact~\cite{contreras2011nciplot, espinosa1999topological, grabowski2001ab,
becke1990simple, silvi1994classification, contreras2011electron,
johnson2010revealing}. It
has been continued by recent work in the framework of partition density
functional theory~\cite{elliott2010partition} with a focus on describing
chemical reactivity~\cite{cohen2006hardness, cohen2007foundations}. Similar to
the methodology presented here, these works also can
describe both covalent and non-covalent interactions between fragments, though
the focus is on an atomic level. The methodology we have presented
here is instead density matrix based, and works at a coarser, fragment level
view. Partitioning is achieved through a fall off of the density matrix in
the linear scaling regime.

In this work, three classical ideas have re-emerged: valence, population
analysis, and bond order (see Mayer\cite{mayer2007bond} for a review). At
the time when these ideas were developed, calculations on systems with even
hundreds of atoms remained out of reach, allowing for careful analysis of
individual atomic contributions. In this work, we have taken those ideas
which were defined at the atomic level of granularity and redefined them for
molecular fragments. By moving to the fragment level, not only is it possible
to derive more chemically meaningful observables, but also to enable
coarse grained analysis of large systems.

In the following few years, the next generation of exascale class supercomputers
promises to enable the routine application of fully quantum mechanical
methods to systems with tens of thousands of atoms. With this, information derived from the
electronic structure will begin to have an impact on entirely new disciplines.
One piece of information from these calculations is the locality of the
electronic structure, which we have used to partition systems and describe
interactions between fragments. The novel fragments generated by this
approach and the graph structures that tie them together are promising
new tools for theoretical studies. Our future work will focus
on applying this methodology to an even wider class of systems, in hopes of
generating novel design rules and insights into large, heterogeneous systems.

\section{Acknowledgements}
This work was supported by the Next-Generation Supercomputer project
(the K computer) and the FLAGSHIP2020 project (Supercomputer Fugaku) within the
priority study5 (Development of new fundamental technologies for high-efficiency
energy  creation, conversion/storage and use) from the Ministry of Education,
Culture, Sports, Science and Technology (MEXT) of Japan.
Experiments presented in this paper were carried out using the Grid'5000
testbed, supported by a scientific interest group hosted by Inria and
including CNRS, RENATER and several Universities as well as other
organizations (see https://www.grid5000.fr).
Calculations were performed using the Hokusai supercomputer system at RIKEN.
This research used resources of the Argonne Leadership Computing Facility, which
is a DOE Office of Science User Facility supported under Contract DE-AC02-06CH11357.
S.M. and L.G. acknowledge support from the European Centre of Excellence MaX, funded within
the Horizon2020 Framework of the European Union under project ID 676598.
L.G. acknowledges support from the EU ExtMOS project and the European Centre of
Excellence EoCoE, funded within the Horizon2020 Framework of the European Union under
project IDs 646176 and 676629, respectively.
L.G., T.N., and W.D. also gratefully acknowledge the joint CEA-RIKEN
collaboration action.
L.E.R. acknowledges support from an EPSRC Early Career Research Fellowship (EP/P033253/1).

We would also like to acknowledge Kazuo Kitaura for his assistance
using the FU program~\cite{fedorov2017modeling} for generating the amino acid
partitioning of the two protein systems and preprocessing of the input
geometry. We would like to thank Thierry Deutsch for valuable discussions and
F\'atima Lucas for providing various test systems and helpful discussions.
Images of molecular systems were generated using VMD~\cite{humphrey1996vmd}.

\appendix
\section{DFT Calculation Details}
\label{sec:DFTdet}
Calculations of each system were performed with the BigDFT
code~\cite{genovese-daubechies-2008}
using density functional theory in the linear scaling
mode~\cite{mohr-daubechies-2014,mohr-accurate-2015} with the
PBE~\cite{perdew1996generalized}
exchange and correlation functional and free boundary conditions.
Hartwigsen-Goedecker-Hutter
(HGH)~\cite{hartwigsen1998relativistic, willand2013norm}
pseudopotentials were used with 11 and 2 valence electrons for copper and
magnesium respectively. Fragmentation and bond order calculations
have been implemented in a new python based pre/post-processing library called
PyBigDFT. These calculations may be run using python notebooks, as have been
included in the supplementary materials along with all geometry files.

In the linear scaling mode of BigDFT, finite distance based cutoffs for kernel
values are employed to maintain the sparsity of the hamiltonian and density
matrix~\cite{mohr2017efficient}. We
note that these distances are much larger than the embedding region sizes
tested in the preceding sections, and as such these a priori cutoffs should
not affect the resulting analysis. One of the key steps for this analysis
is computing the product of the density matrix and overlap matrix. As an extra
precaution, we compute this matrix with no distance cutoff, instead filtering
values of magnitude below $1\times10^{-6}$ using the NTPoly
library~\cite{dawson2018massively}.

\section{Supporting Information}
The python notebooks used to setup calculations, perform analysis, and generate
figures have been included in the supporting information.
\begin{itemize}
  \item \textsc{CR2.ipynb} a python notebook for performing a complexity
  reduction analysis on any of the systems used in this paper.
  \item \textsc{CR2-DoS.ipynb} a python notebook for performing the density of
  states case study.
  \item \textsc{Summary.ipynb} a python notebook for generating summarizing
  figures.
\end{itemize}
In addition to the actual notebooks, static websites generated by these
notebooks have been included (see the \textsc{.html} files in the directory \textsc{Static/}) for each of the system
considered.

Geometry files in the XYZ format, as well as BDA files used to identify amino acid fragments (only of the proteins systems) have also been included in the 
\textsc{Geometries/} and \textsc{BDA/} directories, respectively.

\bibliography{main}

\providecommand{\latin}[1]{#1}
\makeatletter
\providecommand{\doi}
  {\begingroup\let\do\@makeother\dospecials
  \catcode`\{=1 \catcode`\}=2\doi@aux}
\providecommand{\doi@aux}[1]{\endgroup\texttt{#1}}
\makeatother
\providecommand*\mcitethebibliography{\thebibliography}
\csname @ifundefined\endcsname{endmcitethebibliography}
  {\let\endmcitethebibliography\endthebibliography}{}
\begin{mcitethebibliography}{90}
\providecommand*\natexlab[1]{#1}
\providecommand*\mciteSetBstSublistMode[1]{}
\providecommand*\mciteSetBstMaxWidthForm[2]{}
\providecommand*\mciteBstWouldAddEndPuncttrue
  {\def\EndOfBibitem{\unskip.}}
\providecommand*\mciteBstWouldAddEndPunctfalse
  {\let\EndOfBibitem\relax}
\providecommand*\mciteSetBstMidEndSepPunct[3]{}
\providecommand*\mciteSetBstSublistLabelBeginEnd[3]{}
\providecommand*\EndOfBibitem{}
\mciteSetBstSublistMode{f}
\mciteSetBstMaxWidthForm{subitem}{(\alph{mcitesubitemcount})}
\mciteSetBstSublistLabelBeginEnd
  {\mcitemaxwidthsubitemform\space}
  {\relax}
  {\relax}

\bibitem[Hohenberg and Kohn(1964)Hohenberg, and
  Kohn]{hohenberg-inhomogeneous-1964}
Hohenberg,~P.; Kohn,~W. Inhomogeneous Electron Gas. \emph{Phys. Rev.}
  \textbf{1964}, \emph{136}, B864--B871\relax
\mciteBstWouldAddEndPuncttrue
\mciteSetBstMidEndSepPunct{\mcitedefaultmidpunct}
{\mcitedefaultendpunct}{\mcitedefaultseppunct}\relax
\EndOfBibitem
\bibitem[Kohn and Sham(1965)Kohn, and Sham]{kohn-self_consistent-1965}
Kohn,~W.; Sham,~L.~J. Self-Consistent Equations Including Exchange and
  Correlation Effects. \emph{Phys. Rev.} \textbf{1965}, \emph{140},
  A1133--A1138\relax
\mciteBstWouldAddEndPuncttrue
\mciteSetBstMidEndSepPunct{\mcitedefaultmidpunct}
{\mcitedefaultendpunct}{\mcitedefaultseppunct}\relax
\EndOfBibitem
\bibitem[Goedecker(1999)]{goedecker-linear-1999}
Goedecker,~S. {Linear scaling electronic structure methods}. \emph{Rev. Mod.
  Phys.} \textbf{1999}, \emph{71}, 1085--1123\relax
\mciteBstWouldAddEndPuncttrue
\mciteSetBstMidEndSepPunct{\mcitedefaultmidpunct}
{\mcitedefaultendpunct}{\mcitedefaultseppunct}\relax
\EndOfBibitem
\bibitem[Bowler and Miyazaki(2012)Bowler, and Miyazaki]{bowler-O(N)-2012}
Bowler,~D.~R.; Miyazaki,~T. {O(N) methods in electronic structure
  calculations.} \emph{Rep. Prog. Phys.} \textbf{2012}, \emph{75}, 036503\relax
\mciteBstWouldAddEndPuncttrue
\mciteSetBstMidEndSepPunct{\mcitedefaultmidpunct}
{\mcitedefaultendpunct}{\mcitedefaultseppunct}\relax
\EndOfBibitem
\bibitem[Ratcliff \latin{et~al.}(2017)Ratcliff, Mohr, Huhs, Deutsch, Masella,
  and Genovese]{ratcliff-2016-challenges}
Ratcliff,~L.~E.; Mohr,~S.; Huhs,~G.; Deutsch,~T.; Masella,~M.; Genovese,~L.
  Challenges in large scale quantum mechanical calculations. \emph{Wiley
  Interdiscip. Rev.: Comput. Mol. Sci.} \textbf{2017}, \emph{7}, e1290\relax
\mciteBstWouldAddEndPuncttrue
\mciteSetBstMidEndSepPunct{\mcitedefaultmidpunct}
{\mcitedefaultendpunct}{\mcitedefaultseppunct}\relax
\EndOfBibitem
\bibitem[Kitaura \latin{et~al.}(1999)Kitaura, Ikeo, Asada, Nakano, and
  Uebayasi]{kitaura-fragment-1999}
Kitaura,~K.; Ikeo,~E.; Asada,~T.; Nakano,~T.; Uebayasi,~M. Fragment molecular
  orbital method: an approximate computational method for large molecules.
  \emph{Chem. Phys. Lett.} \textbf{1999}, \emph{313}, 701 -- 706\relax
\mciteBstWouldAddEndPuncttrue
\mciteSetBstMidEndSepPunct{\mcitedefaultmidpunct}
{\mcitedefaultendpunct}{\mcitedefaultseppunct}\relax
\EndOfBibitem
\bibitem[Fedorov and Kitaura(2007)Fedorov, and Kitaura]{fedorov-extending-2007}
Fedorov,~D.~G.; Kitaura,~K. Extending the Power of Quantum Chemistry to Large
  Systems with the Fragment Molecular Orbital Method. \emph{J. Phys. Chem. A}
  \textbf{2007}, \emph{111}, 6904--6914\relax
\mciteBstWouldAddEndPuncttrue
\mciteSetBstMidEndSepPunct{\mcitedefaultmidpunct}
{\mcitedefaultendpunct}{\mcitedefaultseppunct}\relax
\EndOfBibitem
\bibitem[Gao(1997)]{gao-toward-1997}
Gao,~J. Toward a Molecular Orbital Derived Empirical Potential for Liquid
  Simulations. \emph{J. Phys. Chem. B} \textbf{1997}, \emph{101},
  657--663\relax
\mciteBstWouldAddEndPuncttrue
\mciteSetBstMidEndSepPunct{\mcitedefaultmidpunct}
{\mcitedefaultendpunct}{\mcitedefaultseppunct}\relax
\EndOfBibitem
\bibitem[Gao(1998)]{gao-a-molecular-1998}
Gao,~J. A molecular-orbital derived polarization potential for liquid water.
  \emph{J. Chem. Phys.} \textbf{1998}, \emph{109}, 2346--2354\relax
\mciteBstWouldAddEndPuncttrue
\mciteSetBstMidEndSepPunct{\mcitedefaultmidpunct}
{\mcitedefaultendpunct}{\mcitedefaultseppunct}\relax
\EndOfBibitem
\bibitem[Wierzchowski \latin{et~al.}(2003)Wierzchowski, Kofke, and
  Gao]{wierzchowski-hydrogen-2003}
Wierzchowski,~S.~J.; Kofke,~D.~A.; Gao,~J. Hydrogen fluoride phase behavior and
  molecular structure: A QM/MM potential model approach. \emph{J. Chem. Phys.}
  \textbf{2003}, \emph{119}, 7365--7371\relax
\mciteBstWouldAddEndPuncttrue
\mciteSetBstMidEndSepPunct{\mcitedefaultmidpunct}
{\mcitedefaultendpunct}{\mcitedefaultseppunct}\relax
\EndOfBibitem
\bibitem[Xie and Gao(2007)Xie, and Gao]{xie-design-2007}
Xie,~W.; Gao,~J. Design of a Next Generation Force Field: The X-POL Potential.
  \emph{J. Chem. Theory Comput.} \textbf{2007}, \emph{3}, 1890--1900\relax
\mciteBstWouldAddEndPuncttrue
\mciteSetBstMidEndSepPunct{\mcitedefaultmidpunct}
{\mcitedefaultendpunct}{\mcitedefaultseppunct}\relax
\EndOfBibitem
\bibitem[Xie \latin{et~al.}(2008)Xie, Song, Truhlar, and
  Gao]{xie-the-variational-2008}
Xie,~W.; Song,~L.; Truhlar,~D.~G.; Gao,~J. The variational explicit
  polarization potential and analytical first derivative of energy: Towards a
  next generation force field. \emph{J. Chem. Phys.} \textbf{2008}, \emph{128},
  234108\relax
\mciteBstWouldAddEndPuncttrue
\mciteSetBstMidEndSepPunct{\mcitedefaultmidpunct}
{\mcitedefaultendpunct}{\mcitedefaultseppunct}\relax
\EndOfBibitem
\bibitem[Xie \latin{et~al.}(2008)Xie, Song, Truhlar, and
  Gao]{xie-incorporation-2008}
Xie,~W.; Song,~L.; Truhlar,~D.~G.; Gao,~J. Incorporation of a QM/MM Buffer Zone
  in the Variational Double Self-Consistent Field Method. \emph{J. Phys. Chem.
  B} \textbf{2008}, \emph{112}, 14124--14131\relax
\mciteBstWouldAddEndPuncttrue
\mciteSetBstMidEndSepPunct{\mcitedefaultmidpunct}
{\mcitedefaultendpunct}{\mcitedefaultseppunct}\relax
\EndOfBibitem
\bibitem[Wang \latin{et~al.}(2012)Wang, Sosa, Cembran, Truhlar, and
  Gao]{wang-multilevel-2012}
Wang,~Y.; Sosa,~C.~P.; Cembran,~A.; Truhlar,~D.~G.; Gao,~J. Multilevel X-Pol: A
  Fragment-Based Method with Mixed Quantum Mechanical Representations of
  Different Fragments. \emph{J. Phys. Chem. B} \textbf{2012}, \emph{116},
  6781--6788\relax
\mciteBstWouldAddEndPuncttrue
\mciteSetBstMidEndSepPunct{\mcitedefaultmidpunct}
{\mcitedefaultendpunct}{\mcitedefaultseppunct}\relax
\EndOfBibitem
\bibitem[Gao and Wang(2012)Gao, and Wang]{gao-variational-2012}
Gao,~J.; Wang,~Y. Communication: Variational many-body expansion: Accounting
  for exchange repulsion, charge delocalization, and dispersion in the
  fragment-based explicit polarization method. \emph{J. Chem. Phys.}
  \textbf{2012}, \emph{136}, 071101\relax
\mciteBstWouldAddEndPuncttrue
\mciteSetBstMidEndSepPunct{\mcitedefaultmidpunct}
{\mcitedefaultendpunct}{\mcitedefaultseppunct}\relax
\EndOfBibitem
\bibitem[Gadre \latin{et~al.}(1994)Gadre, Shirsat, and
  Limaye]{gadre-molecular-1994}
Gadre,~S.~R.; Shirsat,~R.~N.; Limaye,~A.~C. Molecular Tailoring Approach for
  Simulation of Electrostatic Properties. \emph{J. Phys. Chem.} \textbf{1994},
  \emph{98}, 9165--9169\relax
\mciteBstWouldAddEndPuncttrue
\mciteSetBstMidEndSepPunct{\mcitedefaultmidpunct}
{\mcitedefaultendpunct}{\mcitedefaultseppunct}\relax
\EndOfBibitem
\bibitem[Ganesh \latin{et~al.}(2006)Ganesh, Dongare, Balanarayan, and
  Gadre]{ganesh-molecular-2006}
Ganesh,~V.; Dongare,~R.~K.; Balanarayan,~P.; Gadre,~S.~R. Molecular tailoring
  approach for geometry optimization of large molecules: Energy evaluation and
  parallelization strategies. \emph{J. Chem. Phys.} \textbf{2006}, \emph{125},
  104109\relax
\mciteBstWouldAddEndPuncttrue
\mciteSetBstMidEndSepPunct{\mcitedefaultmidpunct}
{\mcitedefaultendpunct}{\mcitedefaultseppunct}\relax
\EndOfBibitem
\bibitem[Gadre and Ganesh(2006)Gadre, and Ganesh]{gadre2006molecular}
Gadre,~S.~R.; Ganesh,~V. Molecular tailoring approach: towards PC-based ab
  initio treatment of large molecules. \emph{J. Theor. Comput. Chem.}
  \textbf{2006}, \emph{5}, 835--855\relax
\mciteBstWouldAddEndPuncttrue
\mciteSetBstMidEndSepPunct{\mcitedefaultmidpunct}
{\mcitedefaultendpunct}{\mcitedefaultseppunct}\relax
\EndOfBibitem
\bibitem[Sahu and Gadre(2014)Sahu, and Gadre]{sahu-molecular-2014}
Sahu,~N.; Gadre,~S.~R. Molecular Tailoring Approach: A Route for ab Initio
  Treatment of Large Clusters. \emph{Acc. Chem. Res.} \textbf{2014}, \emph{47},
  2739--2747\relax
\mciteBstWouldAddEndPuncttrue
\mciteSetBstMidEndSepPunct{\mcitedefaultmidpunct}
{\mcitedefaultendpunct}{\mcitedefaultseppunct}\relax
\EndOfBibitem
\bibitem[Jacob and Neugebauer(2014)Jacob, and Neugebauer]{jacob-subsystem-2014}
Jacob,~C.~R.; Neugebauer,~J. Subsystem density-functional theory. \emph{Wiley
  Interdiscip. Rev.: Comput. Mol. Sci.} \textbf{2014}, \emph{4}, 325--362\relax
\mciteBstWouldAddEndPuncttrue
\mciteSetBstMidEndSepPunct{\mcitedefaultmidpunct}
{\mcitedefaultendpunct}{\mcitedefaultseppunct}\relax
\EndOfBibitem
\bibitem[Krishtal \latin{et~al.}(2015)Krishtal, Sinha, Genova, and
  Pavanello]{krishtal-subsystem-2015}
Krishtal,~A.; Sinha,~D.; Genova,~A.; Pavanello,~M. Subsystem density-functional
  theory as an effective tool for modeling ground and excited states, their
  dynamics and many-body interactions. \emph{J. Phys.: Condens. Matter}
  \textbf{2015}, \emph{27}, 183202\relax
\mciteBstWouldAddEndPuncttrue
\mciteSetBstMidEndSepPunct{\mcitedefaultmidpunct}
{\mcitedefaultendpunct}{\mcitedefaultseppunct}\relax
\EndOfBibitem
\bibitem[Bakowies and Thiel(1996)Bakowies, and Thiel]{bakowies-hybrid-1996}
Bakowies,~D.; Thiel,~W. {Hybrid Models for Combined Quantum Mechanical and
  Molecular Mechanical Approaches}. \emph{J. Phys. Chem.} \textbf{1996},
  \emph{100}, 10580--10594\relax
\mciteBstWouldAddEndPuncttrue
\mciteSetBstMidEndSepPunct{\mcitedefaultmidpunct}
{\mcitedefaultendpunct}{\mcitedefaultseppunct}\relax
\EndOfBibitem
\bibitem[Ferr\'e and \'Angy\'an(2002)Ferr\'e, and \'Angy\'an]{ESPF_2002}
Ferr\'e,~N.; \'Angy\'an,~J.~G. Approximate electrostatic interaction operator
  for QM/MM calculations. \emph{Chem. Phys. Lett.} \textbf{2002}, \emph{356},
  331--339\relax
\mciteBstWouldAddEndPuncttrue
\mciteSetBstMidEndSepPunct{\mcitedefaultmidpunct}
{\mcitedefaultendpunct}{\mcitedefaultseppunct}\relax
\EndOfBibitem
\bibitem[Chung \latin{et~al.}(2015)Chung, Sameera, Ramozzi, Page, Hatanaka,
  Petrova, Harris, Li, Ke, Liu, \latin{et~al.} others]{chung2015oniom}
others,, \latin{et~al.}  The ONIOM method and its applications. \emph{Chem.
  Rev.} \textbf{2015}, \emph{115}, 5678--5796\relax
\mciteBstWouldAddEndPuncttrue
\mciteSetBstMidEndSepPunct{\mcitedefaultmidpunct}
{\mcitedefaultendpunct}{\mcitedefaultseppunct}\relax
\EndOfBibitem
\bibitem[lin(2007)]{lin-QM/MM-2007}
{QM/MM: What have we learned, where are we, and where do we go from here?}
  \emph{Theor Chem. Acc.} \textbf{2007}, \emph{117}, 185\relax
\mciteBstWouldAddEndPuncttrue
\mciteSetBstMidEndSepPunct{\mcitedefaultmidpunct}
{\mcitedefaultendpunct}{\mcitedefaultseppunct}\relax
\EndOfBibitem
\bibitem[Senn and Thiel(2009)Senn, and Thiel]{senn-QM/MM-2009}
Senn,~H.~M.; Thiel,~W. {QM/MM methods for biomolecular systems}. \emph{Angew.
  Chem. Int. Ed.} \textbf{2009}, \emph{48}, 1198--1229\relax
\mciteBstWouldAddEndPuncttrue
\mciteSetBstMidEndSepPunct{\mcitedefaultmidpunct}
{\mcitedefaultendpunct}{\mcitedefaultseppunct}\relax
\EndOfBibitem
\bibitem[Collins and Bettens(2015)Collins, and Bettens]{collins2015energy}
Collins,~M.~A.; Bettens,~R.~P. Energy-based molecular fragmentation methods.
  \emph{Chem. Rev.} \textbf{2015}, \emph{115}, 5607--5642\relax
\mciteBstWouldAddEndPuncttrue
\mciteSetBstMidEndSepPunct{\mcitedefaultmidpunct}
{\mcitedefaultendpunct}{\mcitedefaultseppunct}\relax
\EndOfBibitem
\bibitem[Gordon \latin{et~al.}(2011)Gordon, Fedorov, Pruitt, and
  Slipchenko]{gordon2011fragmentation}
Gordon,~M.~S.; Fedorov,~D.~G.; Pruitt,~S.~R.; Slipchenko,~L.~V. Fragmentation
  methods: A route to accurate calculations on large systems. \emph{Chem. Rev.}
  \textbf{2011}, \emph{112}, 632--672\relax
\mciteBstWouldAddEndPuncttrue
\mciteSetBstMidEndSepPunct{\mcitedefaultmidpunct}
{\mcitedefaultendpunct}{\mcitedefaultseppunct}\relax
\EndOfBibitem
\bibitem[Raghavachari and Saha(2015)Raghavachari, and
  Saha]{raghavachari2015accurate}
Raghavachari,~K.; Saha,~A. Accurate composite and fragment-based quantum
  chemical models for large molecules. \emph{Chem. Rev.} \textbf{2015},
  \emph{115}, 5643--5677\relax
\mciteBstWouldAddEndPuncttrue
\mciteSetBstMidEndSepPunct{\mcitedefaultmidpunct}
{\mcitedefaultendpunct}{\mcitedefaultseppunct}\relax
\EndOfBibitem
\bibitem[He \latin{et~al.}(2014)He, Zhu, Wang, Liu, and Zhang]{he2014fragment}
He,~X.; Zhu,~T.; Wang,~X.; Liu,~J.; Zhang,~J.~Z. Fragment quantum mechanical
  calculation of proteins and its applications. \emph{Acc. Chem. Res.}
  \textbf{2014}, \emph{47}, 2748--2757\relax
\mciteBstWouldAddEndPuncttrue
\mciteSetBstMidEndSepPunct{\mcitedefaultmidpunct}
{\mcitedefaultendpunct}{\mcitedefaultseppunct}\relax
\EndOfBibitem
\bibitem[Li \latin{et~al.}(2014)Li, Li, and Ma]{li2014generalized}
Li,~S.; Li,~W.; Ma,~J. Generalized energy-based fragmentation approach and its
  applications to macromolecules and molecular aggregates. \emph{Acc. Chem.
  Res.} \textbf{2014}, \emph{47}, 2712--2720\relax
\mciteBstWouldAddEndPuncttrue
\mciteSetBstMidEndSepPunct{\mcitedefaultmidpunct}
{\mcitedefaultendpunct}{\mcitedefaultseppunct}\relax
\EndOfBibitem
\bibitem[Hirata \latin{et~al.}(2014)Hirata, Gilliard, He, Li, and
  Sode]{hirata2014ab}
Hirata,~S.; Gilliard,~K.; He,~X.; Li,~J.; Sode,~O. Ab initio molecular crystal
  structures, spectra, and phase diagrams. \emph{Acc. Chem. Res.}
  \textbf{2014}, \emph{47}, 2721--2730\relax
\mciteBstWouldAddEndPuncttrue
\mciteSetBstMidEndSepPunct{\mcitedefaultmidpunct}
{\mcitedefaultendpunct}{\mcitedefaultseppunct}\relax
\EndOfBibitem
\bibitem[Sahu and Gadre(2014)Sahu, and Gadre]{sahu2014molecular}
Sahu,~N.; Gadre,~S.~R. Molecular tailoring approach: a route for ab initio
  treatment of large clusters. \emph{Acc. Chem. Res.} \textbf{2014}, \emph{47},
  2739--2747\relax
\mciteBstWouldAddEndPuncttrue
\mciteSetBstMidEndSepPunct{\mcitedefaultmidpunct}
{\mcitedefaultendpunct}{\mcitedefaultseppunct}\relax
\EndOfBibitem
\bibitem[Pruitt \latin{et~al.}(2014)Pruitt, Bertoni, Brorsen, and
  Gordon]{pruitt2014efficient}
Pruitt,~S.~R.; Bertoni,~C.; Brorsen,~K.~R.; Gordon,~M.~S. Efficient and
  accurate fragmentation methods. \emph{Acc. Chem. Res.} \textbf{2014},
  \emph{47}, 2786--2794\relax
\mciteBstWouldAddEndPuncttrue
\mciteSetBstMidEndSepPunct{\mcitedefaultmidpunct}
{\mcitedefaultendpunct}{\mcitedefaultseppunct}\relax
\EndOfBibitem
\bibitem[Mezey(2014)]{mezey2014fuzzy}
Mezey,~P.~G. Fuzzy electron density fragments in macromolecular quantum
  chemistry, combinatorial quantum chemistry, functional group analysis, and
  shape--activity relations. \emph{Acc. Chem. Res.} \textbf{2014}, \emph{47},
  2821--2827\relax
\mciteBstWouldAddEndPuncttrue
\mciteSetBstMidEndSepPunct{\mcitedefaultmidpunct}
{\mcitedefaultendpunct}{\mcitedefaultseppunct}\relax
\EndOfBibitem
\bibitem[Herbert(2019)]{herbert2019fantasy}
Herbert,~J.~M. Fantasy versus reality in fragment-based quantum chemistry.
  \emph{J. Chem. Phys.} \textbf{2019}, \emph{151}, 170901\relax
\mciteBstWouldAddEndPuncttrue
\mciteSetBstMidEndSepPunct{\mcitedefaultmidpunct}
{\mcitedefaultendpunct}{\mcitedefaultseppunct}\relax
\EndOfBibitem
\bibitem[Mohr \latin{et~al.}(2017)Mohr, Masella, Ratcliff, and
  Genovese]{mohr-fragments1-2017}
Mohr,~S.; Masella,~M.; Ratcliff,~L.~E.; Genovese,~L. Complexity Reduction in
  Large Quantum Systems: Fragment Identification and Population Analysis via a
  Local Optimized Minimal Basis. \emph{J. Chem. Theory Comput.} \textbf{2017},
  \emph{13}, 4079--4088\relax
\mciteBstWouldAddEndPuncttrue
\mciteSetBstMidEndSepPunct{\mcitedefaultmidpunct}
{\mcitedefaultendpunct}{\mcitedefaultseppunct}\relax
\EndOfBibitem
\bibitem[Wiberg(1968)]{wiberg1968application}
Wiberg,~K.~B. Application of the pople-santry-segal CNDO method to the
  cyclopropylcarbinyl and cyclobutyl cation and to bicyclobutane.
  \emph{Tetrahedron} \textbf{1968}, \emph{24}, 1083--1096\relax
\mciteBstWouldAddEndPuncttrue
\mciteSetBstMidEndSepPunct{\mcitedefaultmidpunct}
{\mcitedefaultendpunct}{\mcitedefaultseppunct}\relax
\EndOfBibitem
\bibitem[Mayer(1984)]{mayer1984bond}
Mayer,~I. Bond order and valence: Relations to Mulliken's population analysis.
  \emph{Int. J. Quantum Chem.} \textbf{1984}, \emph{26}, 151--154\relax
\mciteBstWouldAddEndPuncttrue
\mciteSetBstMidEndSepPunct{\mcitedefaultmidpunct}
{\mcitedefaultendpunct}{\mcitedefaultseppunct}\relax
\EndOfBibitem
\bibitem[Borisova and Semenov(1973)Borisova, and
  Semenov]{borisova1973molecular}
Borisova,~N.; Semenov,~S. The molecular-orbital determination of the chemical
  bond order. \emph{Vestn. Leningr. Univ., Ser. 4: Fiz., Khim.} \textbf{1973},
  119--124\relax
\mciteBstWouldAddEndPuncttrue
\mciteSetBstMidEndSepPunct{\mcitedefaultmidpunct}
{\mcitedefaultendpunct}{\mcitedefaultseppunct}\relax
\EndOfBibitem
\bibitem[Armstrong \latin{et~al.}(1973)Armstrong, Perkins, and
  Stewart]{armstrong1973bond}
Armstrong,~D.~R.; Perkins,~P.~G.; Stewart,~J.~J. Bond indices and valency.
  \emph{J. Chem. Soc., Dalton Trans.} \textbf{1973}, 838--840\relax
\mciteBstWouldAddEndPuncttrue
\mciteSetBstMidEndSepPunct{\mcitedefaultmidpunct}
{\mcitedefaultendpunct}{\mcitedefaultseppunct}\relax
\EndOfBibitem
\bibitem[Teeter(1984)]{teeter1984water}
Teeter,~M. Water structure of a hydrophobic protein at atomic resolution:
  Pentagon rings of water molecules in crystals of crambin. \emph{Proc. Natl.
  Acad. Sci. U. S. A.} \textbf{1984}, \emph{81}, 6014--6018\relax
\mciteBstWouldAddEndPuncttrue
\mciteSetBstMidEndSepPunct{\mcitedefaultmidpunct}
{\mcitedefaultendpunct}{\mcitedefaultseppunct}\relax
\EndOfBibitem
\bibitem[Dellafiora \latin{et~al.}(2017)Dellafiora, Galaverna, Reverberi, and
  Dall’Asta]{dellafiora2017degradation}
Dellafiora,~L.; Galaverna,~G.; Reverberi,~M.; Dall’Asta,~C. Degradation of
  aflatoxins by means of laccases from Trametes versicolor: An in silico
  insight. \emph{Toxins} \textbf{2017}, \emph{9}, 17\relax
\mciteBstWouldAddEndPuncttrue
\mciteSetBstMidEndSepPunct{\mcitedefaultmidpunct}
{\mcitedefaultendpunct}{\mcitedefaultseppunct}\relax
\EndOfBibitem
\bibitem[Adamiak \latin{et~al.}(2001)Adamiak, Rypniewski, Milecki, and
  Adamiak]{adamiak20011}
Adamiak,~D.~A.; Rypniewski,~W.~R.; Milecki,~J.; Adamiak,~R.~W. The 1.19 {\AA}
  X-ray structure of 2′-O-Me (CGCGCG) 2 duplex shows dehydrated RNA with
  2-methyl-2, 4-pentanediol in the minor groove. \emph{Nucleic Acids Res.}
  \textbf{2001}, \emph{29}, 4144--4153\relax
\mciteBstWouldAddEndPuncttrue
\mciteSetBstMidEndSepPunct{\mcitedefaultmidpunct}
{\mcitedefaultendpunct}{\mcitedefaultseppunct}\relax
\EndOfBibitem
\bibitem[Dawson and Gygi(2015)Dawson, and Gygi]{dawson2015performance}
Dawson,~W.; Gygi,~F. Performance and accuracy of recursive subspace bisection
  for hybrid DFT calculations in inhomogeneous systems. \emph{J. Chem. Theory
  Comput.} \textbf{2015}, \emph{11}, 4655--4663\relax
\mciteBstWouldAddEndPuncttrue
\mciteSetBstMidEndSepPunct{\mcitedefaultmidpunct}
{\mcitedefaultendpunct}{\mcitedefaultseppunct}\relax
\EndOfBibitem
\bibitem[Mohr \latin{et~al.}(2015)Mohr, Ratcliff, Genovese, Caliste, Boulanger,
  Goedecker, and Deutsch]{mohr-accurate-2015}
Mohr,~S.; Ratcliff,~L.~E.; Genovese,~L.; Caliste,~D.; Boulanger,~P.;
  Goedecker,~S.; Deutsch,~T. Accurate and efficient linear scaling DFT
  calculations with universal applicability. \emph{Phys. Chem. Chem. Phys.}
  \textbf{2015}, \emph{17}, 31360--31370\relax
\mciteBstWouldAddEndPuncttrue
\mciteSetBstMidEndSepPunct{\mcitedefaultmidpunct}
{\mcitedefaultendpunct}{\mcitedefaultseppunct}\relax
\EndOfBibitem
\bibitem[Cole and Hine(2016)Cole, and Hine]{cole2016applications}
Cole,~D.~J.; Hine,~N.~D. Applications of large-scale density functional theory
  in biology. \emph{J. Phys.: Condens. Matter} \textbf{2016}, \emph{28},
  393001\relax
\mciteBstWouldAddEndPuncttrue
\mciteSetBstMidEndSepPunct{\mcitedefaultmidpunct}
{\mcitedefaultendpunct}{\mcitedefaultseppunct}\relax
\EndOfBibitem
\bibitem[Fedorov and Kitaura(2017)Fedorov, and Kitaura]{fedorov2017modeling}
Fedorov,~D.~G.; Kitaura,~K. Modeling and visualization for the fragment
  molecular orbital method with the graphical user interface FU, and analyses
  of protein--ligand binding. In \emph{The fragment molecular orbital method:
  practical applications to large molecular systems}; John Wiley \& Sons, 2017;
  Chapter 3, pp 119--140\relax
\mciteBstWouldAddEndPuncttrue
\mciteSetBstMidEndSepPunct{\mcitedefaultmidpunct}
{\mcitedefaultendpunct}{\mcitedefaultseppunct}\relax
\EndOfBibitem
\bibitem[O'Boyle \latin{et~al.}(2011)O'Boyle, Banck, James, Morley,
  Vandermeersch, and Hutchison]{o2011open}
O'Boyle,~N.~M.; Banck,~M.; James,~C.~A.; Morley,~C.; Vandermeersch,~T.;
  Hutchison,~G.~R. Open Babel: An open chemical toolbox. \emph{J. Cheminf.}
  \textbf{2011}, \emph{3}, 33\relax
\mciteBstWouldAddEndPuncttrue
\mciteSetBstMidEndSepPunct{\mcitedefaultmidpunct}
{\mcitedefaultendpunct}{\mcitedefaultseppunct}\relax
\EndOfBibitem
\bibitem[O'Boyle \latin{et~al.}(2008)O'Boyle, Morley, and
  Hutchison]{o2008pybel}
O'Boyle,~N.~M.; Morley,~C.; Hutchison,~G.~R. Pybel: a Python wrapper for the
  OpenBabel cheminformatics toolkit. \emph{Chem. Cent. J.} \textbf{2008},
  \emph{2}, 1--7\relax
\mciteBstWouldAddEndPuncttrue
\mciteSetBstMidEndSepPunct{\mcitedefaultmidpunct}
{\mcitedefaultendpunct}{\mcitedefaultseppunct}\relax
\EndOfBibitem
\bibitem[Willett(2006)]{willett2006similarity}
Willett,~P. Similarity-based virtual screening using 2D fingerprints.
  \emph{Drug discovery today} \textbf{2006}, \emph{11}, 1046--1053\relax
\mciteBstWouldAddEndPuncttrue
\mciteSetBstMidEndSepPunct{\mcitedefaultmidpunct}
{\mcitedefaultendpunct}{\mcitedefaultseppunct}\relax
\EndOfBibitem
\bibitem[Kamada and Kawai(1989)Kamada, and Kawai]{kamada1989algorithm}
Kamada,~T.; Kawai,~S. An Algorithm for Drawing General Undirected Graphs.
  \emph{Inf. Process. Lett.} \textbf{1989}, \emph{31}, 7–--15\relax
\mciteBstWouldAddEndPuncttrue
\mciteSetBstMidEndSepPunct{\mcitedefaultmidpunct}
{\mcitedefaultendpunct}{\mcitedefaultseppunct}\relax
\EndOfBibitem
\bibitem[Saram{\"a}ki \latin{et~al.}(2007)Saram{\"a}ki, Kivel{\"a}, Onnela,
  Kaski, and Kertesz]{saramaki2007generalizations}
Saram{\"a}ki,~J.; Kivel{\"a},~M.; Onnela,~J.-P.; Kaski,~K.; Kertesz,~J.
  Generalizations of the clustering coefficient to weighted complex networks.
  \emph{Phys. Rev. E} \textbf{2007}, \emph{75}, 027105\relax
\mciteBstWouldAddEndPuncttrue
\mciteSetBstMidEndSepPunct{\mcitedefaultmidpunct}
{\mcitedefaultendpunct}{\mcitedefaultseppunct}\relax
\EndOfBibitem
\bibitem[Estrada(2012)]{estrada2012structure}
Estrada,~E. \emph{The structure of complex networks: theory and applications};
  Oxford University Press, 2012; pp 277--295\relax
\mciteBstWouldAddEndPuncttrue
\mciteSetBstMidEndSepPunct{\mcitedefaultmidpunct}
{\mcitedefaultendpunct}{\mcitedefaultseppunct}\relax
\EndOfBibitem
\bibitem[Vijayabaskar and Vishveshwara(2010)Vijayabaskar, and
  Vishveshwara]{vijayabaskar2010interaction}
Vijayabaskar,~M.; Vishveshwara,~S. Interaction energy based protein structure
  networks. \emph{Biophys. J.} \textbf{2010}, \emph{99}, 3704--3715\relax
\mciteBstWouldAddEndPuncttrue
\mciteSetBstMidEndSepPunct{\mcitedefaultmidpunct}
{\mcitedefaultendpunct}{\mcitedefaultseppunct}\relax
\EndOfBibitem
\bibitem[Fedorov and Kitaura(2007)Fedorov, and Kitaura]{fedorov2007pair}
Fedorov,~D.~G.; Kitaura,~K. Pair interaction energy decomposition analysis.
  \emph{J. Comput. Chem.} \textbf{2007}, \emph{28}, 222--237\relax
\mciteBstWouldAddEndPuncttrue
\mciteSetBstMidEndSepPunct{\mcitedefaultmidpunct}
{\mcitedefaultendpunct}{\mcitedefaultseppunct}\relax
\EndOfBibitem
\bibitem[Sladek \latin{et~al.}(2018)Sladek, Tokiwa, Shimano, and
  Shigeta]{sladek2018protein}
Sladek,~V.; Tokiwa,~H.; Shimano,~H.; Shigeta,~Y. Protein Residue Networks from
  Energetic and Geometric Data: Are They Identical? \emph{J. Chem. Theory
  Comput.} \textbf{2018}, \emph{14}, 6623--6631\relax
\mciteBstWouldAddEndPuncttrue
\mciteSetBstMidEndSepPunct{\mcitedefaultmidpunct}
{\mcitedefaultendpunct}{\mcitedefaultseppunct}\relax
\EndOfBibitem
\bibitem[Barrat and Weigt(2000)Barrat, and Weigt]{barrat2000properties}
Barrat,~A.; Weigt,~M. On the properties of small-world network models.
  \emph{Eur. Phys. J. B} \textbf{2000}, \emph{13}, 547--560\relax
\mciteBstWouldAddEndPuncttrue
\mciteSetBstMidEndSepPunct{\mcitedefaultmidpunct}
{\mcitedefaultendpunct}{\mcitedefaultseppunct}\relax
\EndOfBibitem
\bibitem[Wang \latin{et~al.}(2013)Wang, Liu, Zhang, and
  He]{wang2013electrostatically}
Wang,~X.; Liu,~J.; Zhang,~J.~Z.; He,~X. Electrostatically embedded generalized
  molecular fractionation with conjugate caps method for full quantum
  mechanical calculation of protein energy. \emph{J. Phys. Chem. A}
  \textbf{2013}, \emph{117}, 7149--7161\relax
\mciteBstWouldAddEndPuncttrue
\mciteSetBstMidEndSepPunct{\mcitedefaultmidpunct}
{\mcitedefaultendpunct}{\mcitedefaultseppunct}\relax
\EndOfBibitem
\bibitem[Ganesh \latin{et~al.}(2006)Ganesh, Dongare, Balanarayan, and
  Gadre]{ganesh2006molecular}
Ganesh,~V.; Dongare,~R.~K.; Balanarayan,~P.; Gadre,~S.~R. Molecular tailoring
  approach for geometry optimization of large molecules: Energy evaluation and
  parallelization strategies. \emph{J. Chem. Phys.} \textbf{2006}, \emph{125},
  104109\relax
\mciteBstWouldAddEndPuncttrue
\mciteSetBstMidEndSepPunct{\mcitedefaultmidpunct}
{\mcitedefaultendpunct}{\mcitedefaultseppunct}\relax
\EndOfBibitem
\bibitem[Hua \latin{et~al.}(2010)Hua, Hua, and Li]{hua2010efficient}
Hua,~S.; Hua,~W.; Li,~S. An efficient implementation of the generalized
  energy-based fragmentation approach for general large molecules. \emph{J.
  Phys. Chem. A} \textbf{2010}, \emph{114}, 8126--8134\relax
\mciteBstWouldAddEndPuncttrue
\mciteSetBstMidEndSepPunct{\mcitedefaultmidpunct}
{\mcitedefaultendpunct}{\mcitedefaultseppunct}\relax
\EndOfBibitem
\bibitem[Hua \latin{et~al.}(2013)Hua, Li, and Li]{hua2013generalized}
Hua,~S.; Li,~W.; Li,~S. The Generalized Energy-Based Fragmentation Approach
  with an Improved Fragmentation Scheme: Benchmark Results and Illustrative
  Applications. \emph{ChemPhysChem} \textbf{2013}, \emph{14}, 108--115\relax
\mciteBstWouldAddEndPuncttrue
\mciteSetBstMidEndSepPunct{\mcitedefaultmidpunct}
{\mcitedefaultendpunct}{\mcitedefaultseppunct}\relax
\EndOfBibitem
\bibitem[Steinmann \latin{et~al.}(2012)Steinmann, Ibsen, Hansen, and
  Jensen]{steinmann2012fragit}
Steinmann,~C.; Ibsen,~M.~W.; Hansen,~A.~S.; Jensen,~J.~H. FragIt: a tool to
  prepare input files for fragment based quantum chemical calculations.
  \emph{PLoS One} \textbf{2012}, \emph{7}, e44480\relax
\mciteBstWouldAddEndPuncttrue
\mciteSetBstMidEndSepPunct{\mcitedefaultmidpunct}
{\mcitedefaultendpunct}{\mcitedefaultseppunct}\relax
\EndOfBibitem
\bibitem[Deev and Collins(2005)Deev, and Collins]{deev2005approximate}
Deev,~V.; Collins,~M.~A. Approximate ab initio energies by systematic molecular
  fragmentation. \emph{J. Chem. Phys.} \textbf{2005}, \emph{122}, 154102\relax
\mciteBstWouldAddEndPuncttrue
\mciteSetBstMidEndSepPunct{\mcitedefaultmidpunct}
{\mcitedefaultendpunct}{\mcitedefaultseppunct}\relax
\EndOfBibitem
\bibitem[Le \latin{et~al.}(2012)Le, Tan, Ouyang, and Bettens]{le2012combined}
Le,~H.-A.; Tan,~H.-J.; Ouyang,~J.~F.; Bettens,~R.~P. Combined fragmentation
  method: A simple method for fragmentation of large molecules. \emph{J. Chem.
  Theory Comput.} \textbf{2012}, \emph{8}, 469--478\relax
\mciteBstWouldAddEndPuncttrue
\mciteSetBstMidEndSepPunct{\mcitedefaultmidpunct}
{\mcitedefaultendpunct}{\mcitedefaultseppunct}\relax
\EndOfBibitem
\bibitem[Collins(2012)]{collins2012systematic}
Collins,~M.~A. Systematic fragmentation of large molecules by annihilation.
  \emph{Phys. Chem. Chem. Phys.} \textbf{2012}, \emph{14}, 7744--7751\relax
\mciteBstWouldAddEndPuncttrue
\mciteSetBstMidEndSepPunct{\mcitedefaultmidpunct}
{\mcitedefaultendpunct}{\mcitedefaultseppunct}\relax
\EndOfBibitem
\bibitem[Collins \latin{et~al.}(2014)Collins, Cvitkovic, and
  Bettens]{collins2014combined}
Collins,~M.~A.; Cvitkovic,~M.~W.; Bettens,~R.~P. The combined fragmentation and
  systematic molecular fragmentation methods. \emph{Acc. Chem. Res.}
  \textbf{2014}, \emph{47}, 2776--2785\relax
\mciteBstWouldAddEndPuncttrue
\mciteSetBstMidEndSepPunct{\mcitedefaultmidpunct}
{\mcitedefaultendpunct}{\mcitedefaultseppunct}\relax
\EndOfBibitem
\bibitem[Szalewicz(2012)]{szalewicz2012symmetry}
Szalewicz,~K. Symmetry-adapted perturbation theory of intermolecular forces.
  \emph{Wiley Interdiscip. Rev.: Comput. Mol. Sci.} \textbf{2012}, \emph{2},
  254--272\relax
\mciteBstWouldAddEndPuncttrue
\mciteSetBstMidEndSepPunct{\mcitedefaultmidpunct}
{\mcitedefaultendpunct}{\mcitedefaultseppunct}\relax
\EndOfBibitem
\bibitem[Parrish \latin{et~al.}(2014)Parrish, Parker, and
  Sherrill]{parrish2014chemical}
Parrish,~R.~M.; Parker,~T.~M.; Sherrill,~C.~D. Chemical assignment of
  symmetry-adapted perturbation theory interaction energy components: the
  functional-group SAPT partition. \emph{J. Chem. Theory Comput.}
  \textbf{2014}, \emph{10}, 4417--4431\relax
\mciteBstWouldAddEndPuncttrue
\mciteSetBstMidEndSepPunct{\mcitedefaultmidpunct}
{\mcitedefaultendpunct}{\mcitedefaultseppunct}\relax
\EndOfBibitem
\bibitem[Bader(1991)]{bader1991quantum}
Bader,~R.~F. A quantum theory of molecular structure and its applications.
  \emph{Chem. Rev.} \textbf{1991}, \emph{91}, 893--928\relax
\mciteBstWouldAddEndPuncttrue
\mciteSetBstMidEndSepPunct{\mcitedefaultmidpunct}
{\mcitedefaultendpunct}{\mcitedefaultseppunct}\relax
\EndOfBibitem
\bibitem[Contreras-Garc{\'\i}a \latin{et~al.}(2011)Contreras-Garc{\'\i}a,
  Johnson, Keinan, Chaudret, Piquemal, Beratan, and Yang]{contreras2011nciplot}
Contreras-Garc{\'\i}a,~J.; Johnson,~E.~R.; Keinan,~S.; Chaudret,~R.;
  Piquemal,~J.-P.; Beratan,~D.~N.; Yang,~W. NCIPLOT: a program for plotting
  noncovalent interaction regions. \emph{J. Chem. Theory Comput.}
  \textbf{2011}, \emph{7}, 625--632\relax
\mciteBstWouldAddEndPuncttrue
\mciteSetBstMidEndSepPunct{\mcitedefaultmidpunct}
{\mcitedefaultendpunct}{\mcitedefaultseppunct}\relax
\EndOfBibitem
\bibitem[Espinosa \latin{et~al.}(1999)Espinosa, Souhassou, Lachekar, and
  Lecomte]{espinosa1999topological}
Espinosa,~E.; Souhassou,~M.; Lachekar,~H.; Lecomte,~C. Topological analysis of
  the electron density in hydrogen bonds. \emph{Acta Crystallogr., Sect. B:
  Struct. Sci.} \textbf{1999}, \emph{55}, 563--572\relax
\mciteBstWouldAddEndPuncttrue
\mciteSetBstMidEndSepPunct{\mcitedefaultmidpunct}
{\mcitedefaultendpunct}{\mcitedefaultseppunct}\relax
\EndOfBibitem
\bibitem[Grabowski(2001)]{grabowski2001ab}
Grabowski,~S.~J. Ab initio calculations on conventional and unconventional
  hydrogen bonds study of the hydrogen bond strength. \emph{J. Phys. Chem. A}
  \textbf{2001}, \emph{105}, 10739--10746\relax
\mciteBstWouldAddEndPuncttrue
\mciteSetBstMidEndSepPunct{\mcitedefaultmidpunct}
{\mcitedefaultendpunct}{\mcitedefaultseppunct}\relax
\EndOfBibitem
\bibitem[Becke and Edgecombe(1990)Becke, and Edgecombe]{becke1990simple}
Becke,~A.~D.; Edgecombe,~K.~E. A simple measure of electron localization in
  atomic and molecular systems. \emph{J. Chem. Phys.} \textbf{1990}, \emph{92},
  5397--5403\relax
\mciteBstWouldAddEndPuncttrue
\mciteSetBstMidEndSepPunct{\mcitedefaultmidpunct}
{\mcitedefaultendpunct}{\mcitedefaultseppunct}\relax
\EndOfBibitem
\bibitem[Silvi and Savin(1994)Silvi, and Savin]{silvi1994classification}
Silvi,~B.; Savin,~A. Classification of chemical bonds based on topological
  analysis of electron localization functions. \emph{Nature} \textbf{1994},
  \emph{371}, 683--686\relax
\mciteBstWouldAddEndPuncttrue
\mciteSetBstMidEndSepPunct{\mcitedefaultmidpunct}
{\mcitedefaultendpunct}{\mcitedefaultseppunct}\relax
\EndOfBibitem
\bibitem[Contreras-Garcia and Recio(2011)Contreras-Garcia, and
  Recio]{contreras2011electron}
Contreras-Garcia,~J.; Recio,~J. Electron delocalization and bond formation
  under the ELF framework. \emph{Theor. Chem. Acc.} \textbf{2011}, \emph{128},
  411--418\relax
\mciteBstWouldAddEndPuncttrue
\mciteSetBstMidEndSepPunct{\mcitedefaultmidpunct}
{\mcitedefaultendpunct}{\mcitedefaultseppunct}\relax
\EndOfBibitem
\bibitem[Johnson \latin{et~al.}(2010)Johnson, Keinan, Mori-S{\'a}nchez,
  Contreras-Garc{\'\i}a, Cohen, and Yang]{johnson2010revealing}
Johnson,~E.~R.; Keinan,~S.; Mori-S{\'a}nchez,~P.; Contreras-Garc{\'\i}a,~J.;
  Cohen,~A.~J.; Yang,~W. Revealing noncovalent interactions. \emph{J. Am. Chem.
  Soc.} \textbf{2010}, \emph{132}, 6498--6506\relax
\mciteBstWouldAddEndPuncttrue
\mciteSetBstMidEndSepPunct{\mcitedefaultmidpunct}
{\mcitedefaultendpunct}{\mcitedefaultseppunct}\relax
\EndOfBibitem
\bibitem[Elliott \latin{et~al.}(2010)Elliott, Burke, Cohen, and
  Wasserman]{elliott2010partition}
Elliott,~P.; Burke,~K.; Cohen,~M.~H.; Wasserman,~A. Partition
  density-functional theory. \emph{Phys. Rev. A} \textbf{2010}, \emph{82},
  024501\relax
\mciteBstWouldAddEndPuncttrue
\mciteSetBstMidEndSepPunct{\mcitedefaultmidpunct}
{\mcitedefaultendpunct}{\mcitedefaultseppunct}\relax
\EndOfBibitem
\bibitem[Cohen and Wasserman(2006)Cohen, and Wasserman]{cohen2006hardness}
Cohen,~M.~H.; Wasserman,~A. On hardness and electronegativity equalization in
  chemical reactivity theory. \emph{J. Stat. Phys.} \textbf{2006}, \emph{125},
  1121--1139\relax
\mciteBstWouldAddEndPuncttrue
\mciteSetBstMidEndSepPunct{\mcitedefaultmidpunct}
{\mcitedefaultendpunct}{\mcitedefaultseppunct}\relax
\EndOfBibitem
\bibitem[Cohen and Wasserman(2007)Cohen, and Wasserman]{cohen2007foundations}
Cohen,~M.~H.; Wasserman,~A. On the foundations of chemical reactivity theory.
  \emph{J. Phys. Chem. A} \textbf{2007}, \emph{111}, 2229--2242\relax
\mciteBstWouldAddEndPuncttrue
\mciteSetBstMidEndSepPunct{\mcitedefaultmidpunct}
{\mcitedefaultendpunct}{\mcitedefaultseppunct}\relax
\EndOfBibitem
\bibitem[Mayer(2007)]{mayer2007bond}
Mayer,~I. Bond order and valence indices: A personal account. \emph{J. Comput.
  Chem.} \textbf{2007}, \emph{28}, 204--221\relax
\mciteBstWouldAddEndPuncttrue
\mciteSetBstMidEndSepPunct{\mcitedefaultmidpunct}
{\mcitedefaultendpunct}{\mcitedefaultseppunct}\relax
\EndOfBibitem
\bibitem[Humphrey \latin{et~al.}(1996)Humphrey, Dalke, and
  Schulten]{humphrey1996vmd}
Humphrey,~W.; Dalke,~A.; Schulten,~K. VMD: visual molecular dynamics. \emph{J.
  Mol. Graphics} \textbf{1996}, \emph{14}, 33--38\relax
\mciteBstWouldAddEndPuncttrue
\mciteSetBstMidEndSepPunct{\mcitedefaultmidpunct}
{\mcitedefaultendpunct}{\mcitedefaultseppunct}\relax
\EndOfBibitem
\bibitem[Genovese \latin{et~al.}(2008)Genovese, Neelov, Goedecker, Deutsch,
  Ghasemi, Willand, Caliste, Zilberberg, Rayson, Bergman, and
  Schneider]{genovese-daubechies-2008}
Genovese,~L.; Neelov,~A.; Goedecker,~S.; Deutsch,~T.; Ghasemi,~S.~A.;
  Willand,~A.; Caliste,~D.; Zilberberg,~O.; Rayson,~M.; Bergman,~A.;
  Schneider,~R. {Daubechies wavelets as a basis set for density functional
  pseudopotential calculations.} \emph{J. Chem. Phys.} \textbf{2008},
  \emph{129}, 014109\relax
\mciteBstWouldAddEndPuncttrue
\mciteSetBstMidEndSepPunct{\mcitedefaultmidpunct}
{\mcitedefaultendpunct}{\mcitedefaultseppunct}\relax
\EndOfBibitem
\bibitem[Mohr \latin{et~al.}(2014)Mohr, Ratcliff, Boulanger, Genovese, Caliste,
  Deutsch, and Goedecker]{mohr-daubechies-2014}
Mohr,~S.; Ratcliff,~L.~E.; Boulanger,~P.; Genovese,~L.; Caliste,~D.;
  Deutsch,~T.; Goedecker,~S. Daubechies wavelets for linear scaling density
  functional theory. \emph{J. Chem. Phys.} \textbf{2014}, \emph{140},
  204110\relax
\mciteBstWouldAddEndPuncttrue
\mciteSetBstMidEndSepPunct{\mcitedefaultmidpunct}
{\mcitedefaultendpunct}{\mcitedefaultseppunct}\relax
\EndOfBibitem
\bibitem[Perdew \latin{et~al.}(1996)Perdew, Burke, and
  Ernzerhof]{perdew1996generalized}
Perdew,~J.~P.; Burke,~K.; Ernzerhof,~M. Generalized gradient approximation made
  simple. \emph{Phys. Rev. Lett.} \textbf{1996}, \emph{77}, 3865--3868\relax
\mciteBstWouldAddEndPuncttrue
\mciteSetBstMidEndSepPunct{\mcitedefaultmidpunct}
{\mcitedefaultendpunct}{\mcitedefaultseppunct}\relax
\EndOfBibitem
\bibitem[Hartwigsen \latin{et~al.}(1998)Hartwigsen, G{\oe}decker, and
  Hutter]{hartwigsen1998relativistic}
Hartwigsen,~C.; G{\oe}decker,~S.; Hutter,~J. Relativistic separable dual-space
  Gaussian pseudopotentials from H to Rn. \emph{Phys. Rev. B} \textbf{1998},
  \emph{58}, 3641--3662\relax
\mciteBstWouldAddEndPuncttrue
\mciteSetBstMidEndSepPunct{\mcitedefaultmidpunct}
{\mcitedefaultendpunct}{\mcitedefaultseppunct}\relax
\EndOfBibitem
\bibitem[Willand \latin{et~al.}(2013)Willand, Kvashnin, Genovese,
  V{\'a}zquez-Mayagoitia, Deb, Sadeghi, Deutsch, and
  Goedecker]{willand2013norm}
Willand,~A.; Kvashnin,~Y.~O.; Genovese,~L.; V{\'a}zquez-Mayagoitia,~{\'A}.;
  Deb,~A.~K.; Sadeghi,~A.; Deutsch,~T.; Goedecker,~S. Norm-conserving
  pseudopotentials with chemical accuracy compared to all-electron
  calculations. \emph{J. Chem. Phys.} \textbf{2013}, \emph{138}, 104109\relax
\mciteBstWouldAddEndPuncttrue
\mciteSetBstMidEndSepPunct{\mcitedefaultmidpunct}
{\mcitedefaultendpunct}{\mcitedefaultseppunct}\relax
\EndOfBibitem
\bibitem[Mohr \latin{et~al.}(2017)Mohr, Dawson, Wagner, Caliste, Nakajima, and
  Genovese]{mohr2017efficient}
Mohr,~S.; Dawson,~W.; Wagner,~M.; Caliste,~D.; Nakajima,~T.; Genovese,~L.
  Efficient computation of sparse matrix functions for large-scale electronic
  structure calculations: the CheSS library. \emph{J. Chem. Theory Comput.}
  \textbf{2017}, \emph{13}, 4684--4698\relax
\mciteBstWouldAddEndPuncttrue
\mciteSetBstMidEndSepPunct{\mcitedefaultmidpunct}
{\mcitedefaultendpunct}{\mcitedefaultseppunct}\relax
\EndOfBibitem
\bibitem[Dawson and Nakajima(2018)Dawson, and Nakajima]{dawson2018massively}
Dawson,~W.; Nakajima,~T. Massively parallel sparse matrix function calculations
  with NTPoly. \emph{Comput. Phys. Commun.} \textbf{2018}, \emph{225},
  154--165\relax
\mciteBstWouldAddEndPuncttrue
\mciteSetBstMidEndSepPunct{\mcitedefaultmidpunct}
{\mcitedefaultendpunct}{\mcitedefaultseppunct}\relax
\EndOfBibitem
\end{mcitethebibliography}

\end{document}